\documentclass[prl]{revtex4-1}
\usepackage{graphicx,comment,amsfonts,amsmath}
\usepackage[breaklinks=true,colorlinks=true,linkcolor=blue,urlcolor=blue,citecolor=blue]{hyperref}
\newcommand{\bx}{{\bf r}}
\newcommand{\bxp}{{\bf r}'}
\newcommand{\bk}{{\bf k}}
\newcommand{\be}{{\bf e}}
\newcommand{\bv}{{\bf v}}
\newcommand{\bn}{{\bf \nabla}}

\begin{document}

\title{Computation of rare transitions in the barotropic quasi-geostrophic equations}
\author{Jason Laurie$^{1}$ and Freddy Bouchet$^{2}$}

\affiliation{
$^1$Department of Physics of Complex Systems, Weizmann Institute of Science, Rehovot, 76100, Israel \\
$^2$Laboratoire de Physique, ENS de Lyon, 46, all\'{e}e d'Italie, F69007, Lyon, France}
%\author[cor1]{Jason Laurie}
%\address{Department of Physics of Complex Systems, Weizmann Institute of Science, 234 Herzl Street, Rehovot, 76100, Israel}
%\ead{\mailto{jason.laurie@weizmann.ac.il}}

%\author{Freddy Bouchet}
%\address{Laboratoire de Physique, ENS de Lyon, 46, all\'{e}e d'Italie, F69007, Lyon, France}
%\ead{freddy.bouchet@ens-lyon.fr}

\begin{abstract}
We investigate the theoretical and numerical computation of rare transitions in simple geophysical turbulent models. We consider the barotropic quasi-geostrophic and two-dimensional Navier-Stokes equations in regimes where bistability between two coexisting large-scale attractors exist.  By means of large deviations and instanton theory with the use of an Onsager-Machlup path integral formalism for the transition probability, we show how one can directly compute the most probable transition path between two coexisting attractors analytically in an equilibrium (Langevin) framework and numerically otherwise. We adapt a class of numerical optimization algorithms known as minimum action methods to simple geophysical turbulent models.  We show, that by numerically minimizing an appropriate action functional, in a large deviation limit, one can predict the most likely transition path for a rare transition between two states. By considering examples where theoretical predictions can be made, we show that the minimum action method successfully predicts the most likely transition path. Finally, we discuss the application and extension of such numerical optimization schemes to compute rare transitions observed in direct numerical simulations, experiments and to other, more complex, turbulent systems.  
\end{abstract}

\maketitle

\section{Introduction}

Many turbulent flows related to climate dynamics undergo sporadic random transitions~\cite{benzi_flow_2005}; after long periods of apparent statistical stationarity close to one of the dynamical attractors they spontaneously switch to another dynamical attractor. In recent years, there have been increasing evidence that indicates that the ocean circulation has multiple attractors~\cite{rahmstorf_ocean_2002} corresponding to different regimes of the thermohaline circulation, driven by salinity and temperature differences between the poles and the equator. The transition between such attractors may be related to Dansgaard-Oeschger events~\cite{rahmstorf_ocean_2002,ganopolski_abrupt_2002}. Transitions between two attractors (bistability) is also observed at large scales of ocean currents, for instance for the Kuroshio~\cite{schmeits_bimodal_2001,qiu_kuroshio_2000}.  The importance of possible bistability and abrupt transitions has been emphasized many times including for the planetary atmosphere~\cite{charney_multiple_1979,kravtsov_bimodal_2005,farrell_structural_2003,saravanan_equatorial_1993,chen_sensitivity_2007,weeks_transitions_1997}, where planetary jets may have a huge impact on abrupt climate change~\cite{chavaillaz_southern_2012,toggweiler_midlatitude_2006,farrell_structural_2003}. 

Random transitions in turbulent flows are also extremely prevalent in astrophysics, geophysics, in laboratories and industrial applications. For instance the Earth's magnetic field reversal is a transition between two turbulent attractors just like in magneto-hydrodynamics experiments~\cite{berhanu_magnetic_2007}. Bistability is also observed in two-dimensional turbulence simulations and experiments~\cite{sommeria_experimental_1986,maassen_self-organization_2003,bouchet_random_2009}, Rayleigh-B\'{e}nard convection cells~\cite{chandra_dynamics_2011,niemela_wind_2001,sugiyama_flow_2010, brown_rotations_2006}, and dozens of other three-dimensional fluid flows show this kind of behavior (see for instance~\cite{ravelet_multistability_2004} and~\cite{bouchet_random_2009} for more references). 

Stochastic resonance~\cite{benzi_stochastic_1982,benzi_mechanism_1981} has been advocated as a possible mechanism for the abrupt transitions between glacial and inter-glacial periods, and in relation to bistability of climate dynamics and time varying forces (i.e. the Milankovitch cycles~\cite{hays_variations_1976}). The hypothesis of stochastic resonance is debated~\cite{benzi_stochastic_1982,ghil_cryothermodynamics:_1994} because of the disparity between simple models of only a few degrees of freedom that are used conceptually~\cite{benzi_stochastic_1982,benzi_mechanism_1981}, and more complex models. Stochastic resonance however remains a very interesting possibility. In order to address such issues one should study the attractors and the dynamics of the rare transition between attractors in a hierarchy of models from the simplest to more complex ones used by the climate community. For these complex climate models, that genuinely reproduce the turbulent nature of the Earth's atmosphere and ocean dynamics, such a task is currently inconceivable and seems unreachable is the foreseeable future using direct numerical simulations. The reason is the rarity of the transitions, and the computational complexity of these models.  The main aim of this paper is make a step in the direction of this challenge by studying bistability and the associated transitions in turbulent dynamics using tools that will allow one compute transitions in more complex systems in the near future.

These rare transitions are essential phenomena as they correspond to drastic changes of the complex system behaviour. Moreover, they can not be studied using conventional tools. They contain dynamics occurring on multiple and extremely different timescales, usually with no spectral gap. This prevents the use of classical tools from dynamical system theory. The theoretical understanding of these transitions is an extremely difficult problem due to the complexity, the large number of degrees of freedom, and the non-equilibrium nature of many of these flows.  Up to now, there has been an extremely limited number of theoretical results, where analysis has been limited to analogies with models of very few degrees of freedom~\cite{petrelis_mechanisms_2010}, or to specific classes of systems that can be directly related to equilibrium Langevin dynamics~\cite{bouchet_langevin_2014}. For this reason, the use of non-equilibrium statistical mechanics in order to study those dynamics is necessary. 

The main problem is in how to develop a general theory for these phenomena?
When a complex turbulent flow switches at random from one subregion
of the phase space to another, the first theoretical aim is to characterize
and predict the observed attractors. This is already a non-trivial
task as no picture, based on a potential landscape, is available. Indeed,
this is especially tricky when the transition is not related to any
symmetry breaking. An additional theoretical challenge is in being able to compute the
transition rates between attractors. It is also often the case that
most transition paths from one attractor to another concentrate close
to a single unique path, therefore a natural objective is to compute
this most probable transition path. In order to achieve these goals,
it is convenient to think about the framework of large deviation theory,
in order to describe either, the stationary distribution of the system,
or in computing the transition probabilities of the stochastic process.
In principle, we could argue that from a path integral representation
of the transition probability~\cite{zinn-justin_quantum_2002},
and the study of its semi-classical limit in an asymptotic expansion,
with a well chosen small parameter, we could derive a large deviation
rate function that would characterize the attractors and various other
properties of the system. When this semi-classical approach is relevant, one expects
a large deviation result, similar to that obtained through the
Freidlin-Wentzell theory~\cite{freidlin_random_1984}. If this
notion is correct, then this would explain why these rare transitions
share many analogies with phase transitions in statistical mechanics
and stochastic dynamics with few degrees of freedom.

Therefore with this in mind, we consider the simplest turbulent systems that exhibit random transitions between multiple coexisting attractors. The quasi-geostrophic model with stochastic forces in simple enough to be studied from first principle, in the framework of statistical mechanics and large deviation theory. Moreover this model is relevant to describe some aspects of the largest scales of turbulent geophysical fluid dynamics. The model shares many analogies with the two-dimensional Euler and Navier-Stokes equations~\cite{bouchet_statistical_2012}. These systems include the one-layer quasi-geostrophic model and its subsidiary the two-dimensional Navier-Stokes equations.  For instance, in~\cite{bouchet_random_2009}, the authors observed rare transitions between two quasi-stable large-scale flow configurations, namely a dual-band zonal jet and a vortex dipole for the stochastically forced  two-dimensional Navier-Stokes equations in the limit of weak noise and dissipation. Moreover, multiple zonal jet configurations were observed as dynamical attractors in the quasi-geostrophic model for the same set of parameters~\cite{constantinou_emergence_2013}.  These examples provide the necessary motivation to try and understand rare transitions between two attractors for geophysical fluid flows.  The goal is to develop a theory that will be able to predict the most likely transition path between two attractors without having to resort to direct observations of rare events in Nature, computationally expensive numerical simulations or costly experimental setups. 

To this end, we develop a non-equilibrium statistical mechanics description for the prediction of the most probable rare transitions between two coexisting attractors.  By considering the transition probability of all the possible transition paths between the two states as a Feynman path integral, we apply a saddle-point approximation, in an appropriate limit characterizing the rarity of these transitions, in order to determine the path that yields the greatest contribution to the transition probability.  We decompose the problem into two sub-classes: equilibrium and non-equilibrium.  Through an equilibrium hypothesis, we are able to make direct analytical predictions from the path integral formalism for the most probable transition path.  In this case any transition away from an attractor will become rare in the limit of weak noise.  Alternatively, the non-equilibrium problem is more complex.  In many cases, we are resorted to numerically computing the most probable rare transition through numerical optimization techniques. We outline an appropriate algorithm for use in turbulent models considered here and show that the numerical predictions agree with theoretical results when obtainable.

The layout of this manuscript is as follows: In section~\ref{sec:models} we discuss the class of turbulence models (the barotropic quasi-geostrophic and two-dimensional Navier-Stokes equations) and detail the path integral formalism for the Freidlin and Wentzell (instanton) approach.  In section~\ref{sec:equilibrium}, we overview recent theoretical results of these models in a purely equilibrium Langevin setup.  In such cases, rare trajectories can be directly computed by considering relaxation (deterministic) trajectories of a corresponding dual dynamics. By considering a simple example where a first- and second-order phase transition occurs through a bifurcation of a tri-critical point, we show that the predicted rare trajectories agree with new direct numerical simulations of the system for a transition between two zonal jets. Section~\ref{sec:algorithm} details an numerical optimization algorithm used to compute the most probable rare transition in the barotropic quasi-geostrophic and two-dimensional Navier-Stokes equations in both the equilibrium and non-equilibrium regimes. In section~\ref{sec:numerical_trans} we apply the numerical method from the previous section to several examples of rare transitions in geophysical flows where analytical predictions can be made.  Moreover, we consider an important generalized example of bistability in geophysics: a non-equilibrium rare transition between two distinct zonal jets with topography. We show how the numerical optimization algorithm predicts a rare transition that remains in the set of zonal jet states, thus greatly simplifying the accompanying theory. Finally, we conclude in section~\ref{sec:conclusions} discussing the relevance of the equilibrium and non-equilibrium setups, the advantages and disadvantages of the numerical procedure, and the possible extension of this method to more complex turbulent systems. 

\section{The barotropic quasi-geostrophic and two-dimensional Navier-Stokes equations}
\label{sec:models}
%\subsection{The barotropic quasi-geostrophic model}
%\label{subsec:qg}
 
The most simple turbulent model relevant to bistability in geophysical fluid dynamics is arguably the stochastically forced one-layer barotropic quasi-geostrophic model inside a periodic domain of size $\mathcal{D}=[0, L_x)\times[0,L_y)$ with aspect ratio $\delta = L_x/L_y$:
\begin{subequations}\label{eq:beta-plane}
\begin{align}
 \frac{\partial q}{\partial t}+\mathbf{v}\cdot\mathbf{\nabla}q	&=	-\alpha\omega-\nu\left(-\Delta\right)^{n}\omega+\sqrt{\sigma}\eta,\\
\mathbf{v}=\mathbf{e}_{z}\times\mathbf{\nabla}\psi,&	\qquad	q=\omega+h(\bx)=\Delta\psi+h(\bx),
\end{align}
\end{subequations}
 where $\omega$, $q$, $\bv$ and $\psi$  are the vorticity, potential vorticity, the non-divergent velocity and the streamfunction respectively. Both $\omega$ and $\psi$  are periodic functions in the two spatial directions and are related by $\omega=\Delta\psi$.  The velocity field is recovered using the streamfunction representation that automatically satisfies the incompressibility condition $\nabla\cdot \bv = 0$. The topography if defined through the function $h(\bx)$.  If we set $h\equiv0$, then the barotropic quasi-geostrophic equation~\eqref{eq:beta-plane} reduces to the two-dimensional Navier-Stokes equation with linear friction and hyper-viscosity.  Due to the double periodicity, it is convenient to consider a Fourier representation of the potential vorticity:
  \begin{equation}\label{eq:basis}
q(\bx,t) = \sum_{\bk} q_{\bk}(t)\be_{\bk}(\bx),
 \end{equation}
 where $\be_{\bk}(\bx)=\exp\left(i\bk\cdot\bx\right)/\left(L_xL_y\right)^{1/2}$ is the orthonormal Fourier basis for a doubly periodic domain. 
 We introduce a stochastic noise $\eta$, defined as a sum of random noises
 \begin{equation}\label{eq:forcing}
 \eta(\bx,t) = \sum_{\bk} f_{\bk} \eta_{\bk}(t)\be_{\bk}(\bx),
 \end{equation}
where $\eta_\bk$ are independent, white in time real random noises, such that $\mathbb{E} \left[\eta_\bk(t) \eta_{\bk'}(t')\right] = \delta_{\bk,\bk'}\delta(t-t')$, and $f_{\bk}$ is a complex noise spectrum with randomized phases for each Fourier mode $\bk$.  Consequently, we can define the noise correlation as $\mathbb{E} \left[\eta(\bx,t)\eta(\bxp,t')\right] =C(\bx,\bxp)\delta(t-t')$, where $C(\bx,\bxp)$ the noise correlation matrix can alternatively be represented in Fourier space in terms of the complex noise spectrum: $C(\bx)= \sum_{\bk}C_\bk \be_{\bk}(\bx)= \sum_{\bk} |f_{\bk}|^2 \be_{\bk}(\bx)$.  The noise amplitude $\sigma$ can be associated to the energy injection rate through the normalization of the noise spectrum by 
 \begin{equation}\label{eq:normalization}
\frac{1}{2}\sum_{\bk}\frac{C_{\bk}}{k^{2}}=1,
 \end{equation}
 where $k=|\bk|$ is the wavenumber of the wave vector $\bk$.

Assuming that the topography satisfies the condition $\int_{\mathcal{D}}\, h(\bx)\, d\bx = 0$, the barotropic quasi-geostrophic model~\eqref{eq:beta-plane} on a doubly periodic domain, conserves an infinite number of  dynamical quantities, namely the energy
\begin{equation}\label{eq:energy}
\mathcal{E}[\omega] = -\frac{1}{2}\int_{\mathcal{D}} \, \omega\, \psi \, d\bx,
\end{equation} 
 and an infinite number of Casimirs
 \begin{equation}\label{eq:casimirs}
\mathcal{C}_s[q] = \frac{1}{2}\int_{\mathcal{D}} \, s(q) \, d\bx,
\end{equation} 
where $s(q)$ is any smooth function of the potential vorticity $q$.  On the other hand, a common choice of topography is the beta-plane approximation $h(\bx)=\beta y$, which results in the barotropic quasi-geostrophic model defined upon a beta-plane.  This model is widely used as a simple model for atmospheric and ocean flows, where the curvature of the Earth is approximated by a beta-plane~\cite{pedlosky_geophysical_1987}.  Making the beta-plane approximate results in only the quadratic Casimir $s(q)=q^2/2$ (as well as the energy~\eqref{eq:energy}) being conserved.  In any case, the barotropic quasi-geostrophic model on a beta-plane, the one-layer quasi-geostrophic model, including the two-dimensional Navier-Stokes equations conserve two sign definite quadratic invariants: the energy~\eqref{eq:energy} and the enstrophy $\mathcal{C}_2[q]=(1/2)\int_{\mathcal{D}} \, q^2 \, d\bx$ (where $q=\omega$ for the Navier-Stokes case).  Due to the presence of two quadratic invariants, a simple phenomenological argument~\cite{fjortoft_changes_1953} shows that energy will flow to large scales, while the enstrophy travels towards small scales.  The implications of the inverse energy transfer is of paramount importance in atmospheric and ocean flows.  Restriction imposed by finite sized domains, lead (in the inertial limit) to the condensation of energy at the largest scales, leading to self-organization of the flow into large-scale coherent structures on a background of random turbulent fluctuations.  The explicit form of these structures depends explicitly on the boundary conditions and on the noise correlation.  For periodic boundary conditions, coherent structures in the inertial limit of the Navier-Stokes (Euler) equations can take the form of a vortex dipole or zonal jets~\cite{bouchet_statistical_2012}, whilst only zonal jets are observed in large $\beta$ regimes of the barotropic quasi-geostrophic model~\cite{salmon_equilibrium_1976,bretherton_two-dimensional_1976}.

Using the definition of the energy \eqref{eq:energy}, and the equation of motion~\eqref{eq:beta-plane} we can derive an equation for the energy balance in the system.  By taking the scalar product of \eqref{eq:beta-plane} with the streamfunction $\psi$ and integrating, and applying It\^{o}'s lemma on the noise, we arrive at
\begin{equation}\label{eq:energybalance}
\frac{\partial \mathcal{E}}{\partial t} = -2\alpha\mathcal{E} - \nu \mathcal{H} + \sigma,
\end{equation} 
where $\mathcal H = \int_\mathcal{D} \, \psi(-\Delta)^n \omega  \, d\bx$ corresponds to the dissipation of energy via the hyper-viscosity term.  Assuming the system has achieved a non-equilibrium steady state such that the system reaches an energy balance between the injection $\sigma$ and the dissipation, and by further assuming that the majority of the energy is concentrated at the largest scales (meaning that it is reasonable to neglect energy dissipation through the hyper-viscous term $\mathcal{H}$), we can perform a non-dimensionalization in order to fix the mean energy density to be of order one.   By enforcing that the mean energy density be unity, i.e. that $\mathcal{E}/L^2 = U^2 = L^2/\tau^2 = 1$, where $\tau$ is now the characteristic energy turnover time at the domain scale $L$, from the energy balance equation~\eqref{eq:energybalance} and assuming steady state conditions ($\partial \mathcal{E}/\partial t =0$), we can estimate the typical energy turnover timescale as $\tau = (2\alpha L^4/\sigma)^{1/2}$.  Then by non-dimensionalizing with respect to this timescale, we define new non-dimensional variables as $t'=t\tau$, $\omega' = \omega\tau$, $h'(\bx)=h(\bx)\tau$,
%$\beta' = \beta L\tau$,
 $\alpha' = \alpha\tau$, $\nu' = \nu\tau/ L^{2n}$, and $\eta' = \eta L^2\tau^{1/2}$, resulting in the barotropic model in non-dimensional form:
\begin{subequations}\label{eq:beta-plane2}
\begin{align}
 \frac{\partial q'}{\partial t'}+\mathbf{v}'\cdot\mathbf{\nabla}q'	&=	-\alpha'\omega'-\nu'\left(-\Delta\right)^{n}\omega'+\sqrt{2\alpha'}\eta',\\
\mathbf{v}'=\mathbf{e}_{z}\times\mathbf{\nabla}\psi',&	\qquad	q'=\omega'+h'(\bx) =\Delta\psi'+h'(\bx),
\end{align}
\end{subequations}
%where we have dropped all the primes from the variables.  Here, $\beta'= L/L_{\beta}$ is the ratio of the Rhine's scale $L_{\beta}=1/\beta\tau$ to $L$.  
From this moment on, we will deal with the non-dimensionalized quasi-geostrophic equations~\eqref{eq:beta-plane2} with all primes dropped.

\subsection{Dynamics of the statistically steady state}

For turbulent regimes where the flow is dominated by the presence of large-scale coherent structures, we consider the inertial limits of the barotropic quasi-geostrophic equations~\eqref{eq:beta-plane2}: $\nu < \alpha < 1$.  The type of coherent structures observed are dependent on the topography $h(\bx)$.  Much work has been done in the inertial (Euler) limit, where $h(\bx)=0$.  Here, the main approach has been to advocate the time-scale separation between the inertial dynamic and the slow dissipative and noise dynamics.   Then the invariant measure will concentrate close to the attractors of the two-dimensional Euler equations. A set of attractors can be found using equilibrium statistical mechanics in the form of the Miller-Roberts-Sommeria theory through an energy-Casimir variational problem~\cite{miller_statistical_1990,robert_statistical_1991}. The theory predicts either the formation of a large scale vortex dipole or of a two band zonal jet.  The appearance of the vortex dipole is associated to the degeneracy of the two smallest eigenfunctions of the Laplace operator of the square geometry. For $h(\bx)\neq 0$, the picture is a little more complicated. If we consider the barotropic model on a beta-plane, where $h(\bx)=\beta y$, then the relevant physical parameter that determines the type of coherent structures observed is $\beta$.  For $\beta <1$, we have the usual Euler (or Navier-Stokes) equations, where we expect to observe the invariant measure to concentrate close to vortex dipoles or zonal jets~\cite{bouchet_statistical_2012}.   For rectangular domains, stable parallel flows are formed along the largest scales, where two jets appear in opposite directions.  For $\beta >1$, the $\beta$-effect dominates, which tends to stabilize the parallel flows in the $\be_x$ (zonal) direction (corresponding to the smallest eigenfunction).  Then, possible steady state solutions to the barotropic equations with more than two jets can be observed.  A rough estimate to the number of jets observed can be made by considering the ratio of the domain size $L$ to the Rhine's scale given by $L_{\beta}=1/\beta\tau$.  However, it must be stated that this is only an approximate measure as the structure of the noise correlation will also contribute.  Moreover, many cases of multiple steady state solutions of the barotropic equation with differing number of jets has been observed for the same sets of parameters~\cite{constantinou_emergence_2013}.  These multiple states are assumed to be linearly stable for the unforced (with or without dissipation) dynamics.  Consequently, when one introduces stochastic fluctuations by the addition of a noise, the dynamical attractors become meta-stable, and one may expect to observe rare transitions between several attractors.

\subsection{The large deviation and instanton approach}
\label{subsec:LDtheory}

Transitions between coexisting attractors or the appearance of uncommon large-scale flows are rare events.
There are many ways in which one can study these rare events in general.  However, one of the most promising is the large deviation and instanton approach. This strategy relies on the description of the transition probability of observing a transition between two states in terms of a Feynman path integral derived from the statistical properties of the noise~\cite{feynman_quantum_1965}.  For simplicity, one usually assumes that the system is driven by a white in time noise, however there has been attempts to generalize the formalism to include coloured noises~\cite{wio_path-integral_1989}.  Detailed mathematical derivations of the transition probability can be found in classical textbooks~\cite{freidlin_random_1984,zinn-justin_quantum_2002}. The final result is an Onsager-Machlup path integral~\cite{onsager_fluctuations_1953,machlup_fluctuations_1953} over all possible transition trajectories from a state $q_0$ to a state $q_T$ occurring in time $T$, where each transition is weighted according to some action functional $\mathcal{A}$:
\begin{equation}\label{eq:transprob}
P[q_T,T; q_0,0]\; = \;\int^{q(T)=q_T}_{q(0)=q_0} \; \exp\left(- \frac{\mathcal{A}_{(0,T)}[q]}{2\alpha}\right)\, \mathcal{D}[q].
\end{equation}
Here, deviations from the zero noise (deterministic) relaxation trajectory is represented by a penalty function defined through an action functional $\mathcal{A}_{(0,T)}[q]$.  The action is the time integral of the Lagrangian associated to the dynamical equations:
\begin{equation}\label{eq:action}
\mathcal{A}_{(0,T)}[q] = \int_0^T  \; \mathcal{L}\left[q, \frac{\partial q}{\partial t}\right] \,  dt.
\end{equation}
For the barotropic quasi-geostrophic equations~\eqref{eq:beta-plane2}, the Lagrangian is explicitly 
\begin{align}\label{eq:Lagrangian}
\mathcal{L}\left[q, \frac{\partial q}{\partial t} \right] &= \frac{1}{2} \int_{\mathcal{D}} \int_{\mathcal{D}}\left[\frac{\partial q}{\partial t} + \bv\cdot\nabla q + \alpha \omega + \nu(-\Delta)^n \omega \right](\bx)\nonumber
\\&\times C^{-1}(\bx,\bx') \left[\frac{\partial q}{\partial t} + \bv\cdot\nabla q + \alpha \omega + \nu(-\Delta)^n \omega \right](\bx') \, d\bx \, d\bx',
\end{align}
where $C^{-1}(\bx,\bxp)$ is the formal inverse of the noise correlation, such that $\int_{\mathcal{D}}\, C(\bx,\bx_1)C^{-1}(\bx_1,\bx')\, d{\bx_1} = \delta(\bx - \bx')$.    

For rare probabilities, the path integral is a Laplace integral and one can often perform a saddle-point approximation around the global minimum of the action functional to get a leading order approximation to the transition probability.  This estimate will be based on the action of the trajectory that globally minimizes the action functional.  This global minimizer will be the most probable transition path going from state $q_0$ to $q_T$ in time $T$. At its most simple, it will consist of the most probable fluctuation path out of the initial attractor to the edge of the basin of attraction of a neighouring attractor (known as an instanton), and then the relaxation to the second attractor. Mathematically, this is defined as
\begin{equation}\label{eq:instanton}
q^*(t) =\underset{\left.\left\{q \; \right| \; q(\bx,0) = q_0, \;q(\bx,T)=q_T \right\}}{\operatorname{arg~min}} \; \mathcal{A}_{(0,T)}[q].
\end{equation}
When the saddle-point approximation is valid, the transitions are rare and are clustered around the instanton path.  
As global minima, the most probable paths are critical points of the action functional~\eqref{eq:action}, and satisfy the corresponding Euler-Lagrange equations 
\begin{equation}\label{eq:Euler-Lagrange_general}
\frac{\delta\mathcal{L}}{\delta q} = \frac{d}{dt}\frac{\delta \mathcal{L}}{\delta \dot{q}},
\end{equation}
where $\dot{q}=\partial q/\partial t$. The Euler-Lagrange equations~\eqref{eq:Euler-Lagrange_general} can be re-expressed in terms of an instanton Hamiltonian $\mathcal{H}[q,p]$ for canonical variables $q$ and $p=\delta\mathcal{L}/\delta \dot{q}$:
\begin{subequations}
\begin{align}
\frac{\partial q}{\partial t} & = \frac{\delta \mathcal{H}}{\delta p},\\
\frac{\partial p}{\partial t} & = -\frac{\delta \mathcal{H}}{\delta q}.
\end{align} 
\end{subequations}
The instanton Hamiltonian should not be confused with the energy $\mathcal{E}$.  The instanton Hamiltonian is a quantity that remains conserved by the dynamics of the most probable transition path (instanton and relaxation).  Therefore, $\mathcal{H}$ becomes an extremely useful quantity for numerical purposes where it can be used to verify whether a transition path is a critical point of the action functional~\eqref{eq:action}.  For the quasi-geostrophic dynamics~\eqref{eq:beta-plane2} the instanton Hamiltonian $\mathcal{H}$ is given by
\begin{align}\label{eq:instantonHam}
\mathcal{H}[q,p] &=  \frac{1}{2}\int \; p(\bx)C(\bx,\bxp)p(\bxp) \, d\bx \;d\bxp  - \int  \; p(\bx)\left[\bv \cdot \bn q + \;\alpha \omega +\;\nu \left(-\Delta^n\right) \omega \right](\bx)\, d\bx.
\end{align}
Then, explicitly the Euler-Lagrange equations become
\begin{subequations}\label{eq:Euler-Lagrange}
\begin{align}
\frac{\partial p}{\partial t}+\bv\cdot\bn p&=\Delta^{-1}(\be_{z}\cdot[\bn q\times\bn p])+\alpha p+\nu\left(-\Delta\right)^{n}p,\\
\frac{\partial q}{\partial t}+\mathbf{v}\cdot\nabla q&=-\alpha\omega-\nu\left(-\Delta\right)^{n}\omega+\int_{\mathcal{D}}\; C({\bx},\bx')p(\bx')\, d\bx',
\end{align}
\end{subequations}
subject to the boundary conditions $q(\bx,0)=q_0$ and $q(\bx, T)=q_T$.  The Euler-Lagrange equations~\eqref{eq:Euler-Lagrange} are also known as the instanton equations as the most probable transition path will satisfy them.  It should be make clear that any transition path that is a critical point of the action functional may satisfy the Euler-Lagrange equations~\eqref{eq:Euler-Lagrange}, not only the most probable one~\eqref{eq:instanton}. On general grounds, one should expect multiple solutions to the instanton equations, hence leading one to compare their respective action values in order to determine the most likely path.  It should be emphasized that Eqs.~\ref{eq:Euler-Lagrange} are only valid in the Freidlin-Wentzell limit of the transition being rare, where the saddle-point approximation is valid.

%paragraph about dns and why this is not idea lthen Forward flux sampling to compute transition rates
A straightforward study of rare transitions, through direct numerical simulation of the governing equations is nearly always impracticable. This is mainly a complexity problem, due to the large number of degrees of freedom involved for genuine turbulent flows, and the extremely long time between two successive transitions.   The path integral approach provides a way to systematically determine the most likely transition path between two attractors.  Through the action functional $\mathcal{A}$ with the {\it a priori} given attractors one can predict the most probable rare transition by considering the local action minimizers or by solving the instanton equations with appropriate boundary conditions.  Theoretically, this problem is also extremely difficult, and for turbulent flows, can only be achieved in the simplest of circumstances (see section~\ref{sec:numerical_trans}).  Alternatively one can resort to numerical approaches.  Numerical algorithms exist that compute the most probable transition paths by iteratively converging towards local action minimizers (see section~\ref{sec:algorithm}), or by directly solving the boundary value problem associated to the instanton equations~\eqref{eq:Euler-Lagrange} (for instance see~\cite{grafke_arclength_2014}).

In the cases where the dynamics are in equilibrium, i.e. that they satisfy Langevin dynamics, every transition out of an attractor can be rare.  This allows for the direct computation of rare transitions through deterministic relaxation trajectories in a related dual system~\cite{bouchet_langevin_2014}, as will be discussed in the next section.

\section{Equilibrium Langevin dynamics of the two-dimensional quasi-geostrophic equations}
\label{sec:equilibrium}

This section contains a brief overview of recent theoretical work~\cite{bouchet_langevin_2014} on the Langevin dynamics of the quasi-geostrophic equations.  We introduce the Langevin dynamics description for the quasi-geostrophic equations, where a specific relationship between the noise correlation and the kernel of the potential force invokes the Langevin property.  By specifying a particular structure of the potential term, we consider a Langevin dynamics with a first-order phase transition between coexisting zonal flow attractors.  Through the equilibrium Langevin dynamics theory, we can analytically predict the most probable rare transition paths between two attractors by considering an effective potential landscape and relaxation (unforced) trajectories of a dual dynamics.  This also yields an Arrhenius law for the transition probability. The final part of this section is dedicated to the direct numerical simulations of the bistable system considered in~\cite{bouchet_langevin_2014}, verifying numerically that the theoretical predictions hold.

\subsection{The two-dimensional quasi-geostrophic Langevin equations\label{sec:The-2D-Euler-QG-dynamics}}

The Langevin formalism was previously considered for the two-dimensional quasi-geostrophic and Euler equations in~\cite{bouchet_langevin_2014}.  We explain why the two main hypotheses of Langevin
dynamics (Liouville property and conservation of the potential related
to the transversality condition) are verified when the
kernel in front of the gradient part and the noise autocorrelation
are identical.

The Langevin dynamics associated to the quasi-geostrophic
equations in a periodic domain $\mathcal{D}=[0,2\pi\delta)\times[0,2\pi)$
with aspect ratio $\delta$ given by
\begin{subequations}\label{eq:qg_langevin}
\begin{align}
\frac{\partial q}{\partial t}+\mathbf{v}\left[q-h\right]\cdot\mathbf{\nabla}q & =  -\alpha\int_{\mathcal{D}}\, C({\bf r},{\bf r}')\frac{\delta\mathcal{G}}{\delta q({\bf r}')}\,d{\bf r}'+\sqrt{2\alpha\gamma}\eta,\\
\mathbf{v}=\mathbf{e}_{z}\times\mathbf{\nabla}\psi,\quad \omega&=\Delta\psi, \quad q=\omega+h(\mathbf{r}),
\end{align}
\end{subequations}
with potential $\mathcal{G}$. The stochastic force $\eta$ is a Gaussian
process, white in time, with correlation function $\mathbb{E}\left[\eta(\mathbf{r},t)\eta(\mathbf{r}',t')\right]=C(\mathbf{r},\mathbf{r}')\delta(t-t')$.  The topography $h(\mathbf{r})$ is such that $\int_{\mathcal{D}}\, h\left(\mathbf{r}\right)\, d{\bf r}=0$ and we also introduce a new parameter $\gamma$ that will control the relative strength of the noise to the potential term.   For the Langevin description to be correct, the potential $\mathcal{G}$ must consist of conserved quantities of the inviscid ($\alpha=0$) dynamics of~\eqref{eq:beta-plane2}.  Moreover, the deterministic equations for Langevin dynamics (Eqs. (\ref{eq:qg_langevin})
for $\alpha =0$) essentially correspond to a transport equation by a non-divergent velocity field, leading to a Liouville property for the nonlinear advection term $\bv\cdot\nabla q$.  A more detailed discussion of the Langevin assumptions and results can be found in~\cite{bouchet_langevin_2014}.  

%We consider $G$ to be the Green's function of the Laplacian operator ($G=\Delta^{-1}$)
%for doubly periodic functions with zero averages. Then, the equations relating  the potential vorticity $q$, the stream function $\psi$,
%and the velocity are inverted as
%\begin{equation}
%\psi(\mathbf{r})=\int_{\mathcal{D}}\, G\left(\mathbf{r},{\bf r}'\right)\left[q({\bf r}')-h({\bf r}')\right]\, d{\bf r}',
%\end{equation}
%and 
%\begin{equation}
%\mathbf{v}\left[\omega\right](\mathbf{r})=\int_{\mathcal{D}}\,\mathbf{e}_{z}\times\nabla_{\mathbf{r}'}G\left(\mathbf{r},{\bf r}'\right)\omega(\mathbf{r}')\, d{\bf r}',\label{eq:inversion-velocity}
%\end{equation}
%respectively. Here, $\mathbf{v}\left[\omega\right]$ is the operator that
%allows us to compute the velocity from the vorticity. When $h=0$, these dynamics correspond to the two-dimensional Euler equilibrium dynamics.

%\subsection{Conserved quantity and Liouville property\label{sub:Conserved-quantity-and-Liouville}}
As with the quasi-geostrophic equations~\eqref{eq:beta-plane2}, the equilibrium quasi-geostrophic dynamics conserve the energy $\mathcal{E}$ and an infinite number of Casimirs $\mathcal{C}_s$ given by Eqs.~\eqref{eq:energy} and~\eqref{eq:casimirs} respectively for the deterministic ($\alpha=0$) dynamics.
%From the velocity-vorticity relationship, it is easily checked that the kinetic energy can be expressed as 
%\begin{equation}
%\mathcal{E}=-\frac{1}{2}\int_{\mathcal{D}}\,\left[q-h\left(\mathbf{r}\right)\right]\psi\, d\mathbf{r}=\frac{1}{2}\int_{\mathcal{D}}\,\left(\nabla\psi\right)^{2}\, d\mathbf{r},\label{eq:Energy-QG}
%\end{equation}
%and, for any sufficiently smooth real function $s$, the Casimir functionals are defined as
%\begin{equation}
%\mathcal{C}_{s}=\int_{\mathcal{D}}\, s(q)\, d\mathbf{r},
%\end{equation}
%which are all conserved quantities of the deterministic quasi-geostrophic dynamics (Eqs. (\ref{eq:beta-plane}) for $\alpha=0$). For any $s$,
%and any $\beta$ the functional 
Then, the correct choice of the potential $\mathcal{G}$ for Langevin dynamics will consist of a combination of these conserved quantities:
\begin{equation}
\mathcal{G}=\mathcal{C}_{s}+\beta\mathcal{E}.
\end{equation}

\subsection{Reversed dynamics and the relaxation equation}
\label{sub:Instanton-equation-QG}

For the two-dimensional Euler or quasi-geostrophic equations, the time-reversed dynamics defined as $q_r(t)=I[q(T-t)]$ also satisfies a Langevin dynamics through a set of symmetries with the relevant involution operator $I[\cdot]$
corresponding to a time reversal being
\begin{equation}
I\left[q\right]=-q.
\end{equation}
Then, the dual process is given by (\ref{eq:qg_langevin}) but with $\mathbf{v}\left[q-h\right]\cdot\nabla q\to \mathbf{v}\left[q+h\right]\cdot\nabla q$ and $\mathcal{G}\left[q\right]\to\mathcal{G}\left[-q\right]$ giving
\begin{subequations}\label{eq:qg_dual}
\begin{align}
\frac{\partial q}{\partial t}+\mathbf{v}\left[q+h\right]\cdot\mathbf{\nabla}q & =  -\alpha\int_{\mathcal{D}}\, C({\bf r},{\bf r}')\frac{\delta\mathcal{G}[-q]}{\delta q({\bf r}')}\,d{\bf r}'+\sqrt{2\alpha\gamma}\eta,\\
\mathbf{v}=\mathbf{e}_{z}\times\mathbf{\nabla}\psi,\quad \omega&=\Delta\psi, \quad q=\omega-h(\mathbf{r}),
\end{align}
\end{subequations}

We observe that for the two-dimensional Euler equations ($h=0$), the dual dynamics~\eqref{eq:qg_dual} agree with the original dynamics~\eqref{eq:qg_langevin} if the potential $\mathcal{G}$ is even. Then we conclude that the dynamics
are time-reversible and detailed balance is verified. If however, $\mathcal{G}$ is not even or that $h\neq 0$, then the dynamics are not time-reversible and the original dynamics are conjugated
to another Langevin dynamics where $h$ has to be replaced by $-h$ and $\mathcal{G}$ by $\mathcal{G}\left[-q\right]$. In this case, detailed balance is not verified.

For Langevin dynamics, the instantons
from one attractor to a saddle are given by the reverse of the relaxation
paths of the corresponding dual dynamics. The relaxation paths are simply the deterministic trajectories of the Langevin dynamics~\eqref{eq:qg_langevin} with $\gamma=0$.  Therefore, for the barotropic quasi-geostrophic equations the equation for the relaxation paths are
\begin{equation}
\frac{\partial q}{\partial t}+\mathbf{v}\left[q+h\right]\cdot\nabla q=-\alpha\int_{\mathcal{D}}\, C({\bf r},{\bf r}')\frac{\delta\mathcal{G}}{\delta q({\bf r}')}\left[-q\right] d{\bf r}'.\label{eq:relaxation-paths-QG}
\end{equation}
Eq.~\eqref{eq:relaxation-paths-QG} are known as the relaxation equations.  They provide the means to directly compute, through deterministic means, the instanton trajectories from an attractor to a saddle by considering the relaxation paths, defined by \eqref{eq:relaxation-paths-QG} from the saddle to the attractor.

\subsection{The Energy-enstrophy ensemble and physical dissipation}
\label{sub:Energy-enstrophy-physicaldiscussion}

A special case of Langevin dynamics are when the potential is given by the following form 
\begin{equation}
\mathcal{G}=\int_{\mathcal{D}}\,\frac{q^{2}}{2}\,d\mathbf{r}+\beta\mathcal{E}.\label{eq:Energy-Enstrophy}
\end{equation}
This structure is referred to as the potential enstrophy ensemble (when $\beta=0$), the
enstrophy ensemble (when $\beta=0$ and $h=0$), or generally as the energy-enstrophy
ensemble. The properties of the corresponding invariant measures have been discussed on a number of occasions, starting with the works of Kraichnan \cite{kraichnan_two-dimensional_1980}
in the case of Galerkin truncations of the dynamics, and for some cases without
discretization, see for instance \cite{Bouchet_invariant_2010}
and references therein. 

For specific choices of the potential $\mathcal{G}$ and of the noise correlation
$C$, the friction term can also be identified with a classical physical
dissipation mechanism. For instance, if $C({\bf r},{\bf r}')=\Delta\delta(\mathbf{r}-\mathbf{r}')$,
and the potential takes the form of (\ref{eq:Energy-Enstrophy}), then
the dissipative term on the right hand side of (\ref{eq:qg_langevin})
is 
\begin{equation}
-\alpha\int_{\mathcal{D}}\, C({\bf r},{\bf r}')\frac{\delta\mathcal{G}}{\delta q({\bf r}')}\left[q\right]\,d{\bf r}'=\alpha\Delta q-\alpha\beta q,
\end{equation}
which leads to a diffusion type dissipation with viscosity $\alpha$ and
a linear friction with friction parameter $\alpha\beta$. Such a linear
friction can model the effects of three-dimensional
boundary layers on the quasi two-dimensional bulk vorticity, that appear in
experiments with a very large aspect ratio, rotating tank experiments,
or soap film experiments.

The fact that for the enstrophy ensemble, the quasi-potential is simply the
enstrophy, the relaxation and fluctuation paths can be easily
computed explicitly in many scenarios, as is discussed in \cite{bouchet_statistical_2012}.

For the majority of the other cases, the dissipative term on the right hand
side of (\ref{eq:qg_langevin}) cannot be identified as a microscopic
dissipation mechanism nor as a physical mechanism.  There is however another
possible interpretation of this kind of friction term. As explained
in \cite{bouchet_simpler_2008}, entropy maxima subjected to constraints
related to the conservation of energy and the distribution of vorticity,
are also extrema of energy-Casimir functionals.
By analogy with the Allen-Cahn equation in statistical mechanics,
that uses the free energy as a potential, it seems reasonable to describe
the largest scales of turbulent flows as evolving through a gradient
term of the energy-Casimir functional. Such models have been considered
in the past (see, for example \cite{chavanis_generalized_2003,chavanis_dynamical_2009}
and references therein). At this stage, this should be considered
as a phenomenological approach, as no
clear theoretical results exist to support this view.

\subsection{Phase transitions and instantons between zonal flows in the equilibrium quasi-geostrophic equations}
\label{sub:Phase-transition-QG}

In order to fully determine the quasi-geostrophic Langevin dynamics
(\ref{eq:qg_langevin}), we need to specify the topography function
and the potential $\mathcal{G}$. Given the infinite number of conserved quantities for the quasi-geostrophic dynamics, there
are many possible choices. We are interested in the description of the phenomenology
of phase transitions and instanton theory in situations of first-order transitions.  Therefore, we will illustrate such a phenomenology through
an example originally discussed in~\cite{bouchet_langevin_2014}. 

We begin by choosing a topography given by $h\left(\mathbf{r}\right)=H\cos\left(2y\right)$ on a periodic domain $[0,2\pi\delta)\times[0,2\pi)$,
such that 
\begin{equation}
q=\Delta\psi+H\cos\left(2y\right),
\end{equation}
and consider the potential 
\begin{equation}
\mathcal{G}=\mathcal{C}+\beta\mathcal{E},\label{eq:Potential-Zonal-Transitions-1}
\end{equation}
with energy $\mathcal{E}$ (Eq.~\eqref{eq:energy}), with $\beta$ the inverse temperature
and where $\mathcal{C}$ is the Casimir functional 
\begin{equation}
\mathcal{C}=\int_{\mathcal{D}}\,\frac{q^{2}}{2}-\epsilon\frac{q^{4}}{4}+a_{6}\frac{q^{6}}{6}\, {\rm d}\mathbf{r},\label{eq:C-a4}
\end{equation}
where we assume that $a_{6}>0$ and $\beta= -1+\epsilon$.

In~\cite{bouchet_langevin_2014}, it was shown that for the potential \eqref{eq:Potential-Zonal-Transitions-1}, with small $\beta$  and $\epsilon>0$ we expect to observe a first order phase transition.  When $H=0$, a bifurcation occurs for $\beta=-1$ ($\epsilon=0$) which can be easily checked (see {\cite{corvellec_complete_2012}).
This bifurcation is due to the vanishing of the Hessian
at $\beta=-1$ ($\epsilon=0$). As discussed in many
papers \cite{chavanis_classification_1996,venaille_statistical_2009,bouchet_statistical_2012,corvellec_complete_2012},
for the quadratic Casimir functional $\mathcal{C}_{2}=\int_{\mathcal{D}}\,q^{2}/2\, {\rm d}{\bf r}$,
the first bifurcation involves the eigenfunction of $-\Delta$ with
the lowest eigenvalue. If we assume that the aspect ratio $\delta<1$,
then the smallest eigenvalue is the one corresponding to the zonal
mode proportional to $\cos\left(y\right)$. Because we are interested by transitions
between two zonal states, we assume from now on that $\delta<1$.

For non-zero, but sufficiently small, $H$ there will still be a bifurcation
for $\epsilon$ close to zero. This is the regime that we wish to consider. The null space of the Hessian is spanned by eigenfunctions
$\cos\left(y\right)$ and $\sin\left(y\right)$, therefore as a consequence, for small enough $\epsilon$ and $H$, we expect that the bifurcation can be described by a normal
form involving only the projection of the field $q$ onto the null
space.   It was shown in~\cite{bouchet_langevin_2014}, that by tackling the problem perturbatively, assuming that $\epsilon\ll 1$, $H^2\ll 1$ and $a_6H^2\ll \epsilon$, the Langevin dynamics can be described by an effective potential given by

\begin{equation}
\mathcal{G}=\delta \pi^{2} G(A,B)\label{eq:Relation-G-GAB}
\end{equation}
with $G$ given at leading order by 
\begin{align}
G(A,B)=&-\frac{{H^{2}}}{3}+\left(\epsilon-\frac{\epsilon H^{2}}{6}+\frac{5a_{6}H^{4}}{216}\right)\left(A^{2}+B^{2}\right)\nonumber\\
&+\left(-\frac{3\epsilon}{8}+\frac{25a_{6}H^{2}}{144}\right)\left(A^{2}+B^{2}\right)^{2}+\frac{5a_{6}}{24}\left(A^{2}+B^{2}\right)^{3}+\frac{5a_{6}H^{2}}{72}\left(A^{2}-B^{2}\right)^{2},\label{eq:critical_points}
\end{align}
around the potential vorticity field of
\begin{equation}
q =  - \frac{H}{3}\cos(2y) -A\cos(y)-B\sin(y) + \mathcal{O}\left(\epsilon\right)  + \mathcal{O}\left(\epsilon A^2\right) +  \mathcal{O}\left(a_6A^4\right).
\end{equation}

The fact that $G$ is a normal form for small enough $\epsilon$, $a_{6}$,
and $H$, implies that the gradient of $\mathcal{G}$ in the directions
transverse to $q=-A\cos\left(y\right)-B\sin\left(y\right)$ are much steeper than the gradient
of $G$. 

\begin{figure}
\begin{center}
\includegraphics[width=0.56\columnwidth]{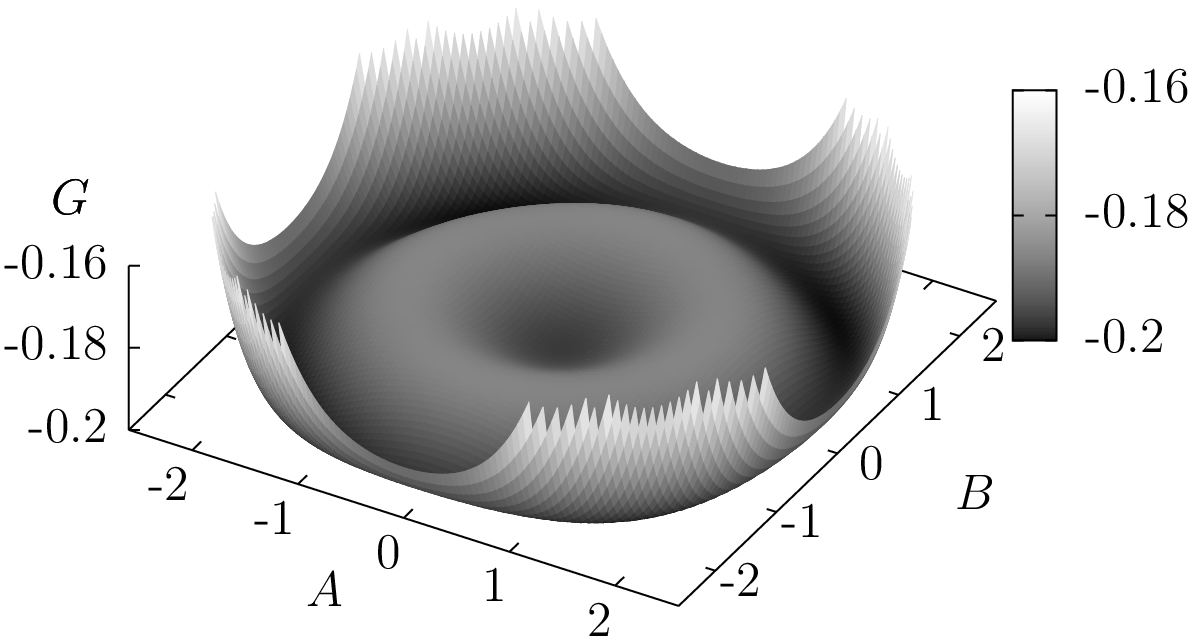}
\hfill
\includegraphics[width=0.4\columnwidth]{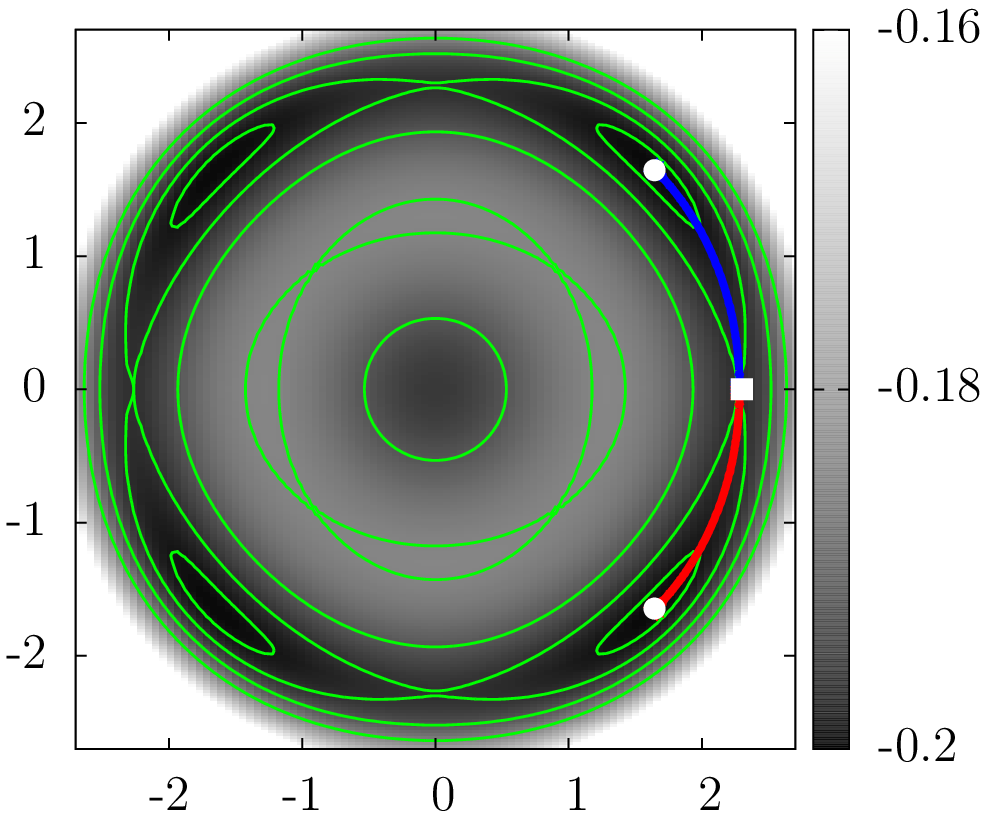}
\caption{\label{fig:G-A-B} Figure adapted from~\cite{bouchet_langevin_2014}. Surface plot (left) and contour plot (right) of the reduced potential surface $G(A,B)$ (see Eq.
(\ref{eq:critical_points})) for parameters: $\epsilon=1.6\times10^{-2}$, $H=7.746\times 10^{-1}$, $a_{6}=2.6\times10^{-3}$. For these parameter, $G$ has four global minima with
$\left|A\right|=\left|B\right|$ and one local minima at $A=B=0$.  This structure with four non-trivial attractors
is due to symmetry breaking imposed by the topography $h(y)=H\cos\left(2y\right)$.  Level contours are shown in green.  The most probable transition path is shown by the red and blue curves. The instanton (red curve) is the reverse trajectory of a relaxation path from the saddle (white square) to the attractor (white circle).}
\end{center}
\end{figure}

We observe that the term proportional to $\left(A^{2}-B^{2}\right)^{2}$ breaks the
symmetry between $A$ and $B$. Its minimization imposes that $A^{2}=B^{2}$.
Then either $A=B$, or $A=-B$. If we take into account that minimizing
with respect to $A^{2}+B^{2}$ will give only the absolute value
of $A$, we can surmise that we will have four equivalent non-trivial solutions: 
\begin{equation}\label{eq:attractors}
q_{i}=-\frac{H}{3}\cos\left(2y\right)+\sqrt{2}\left|A\right|(\epsilon,a_{6})\cos(y+\phi_{i}),
\end{equation}
with $\phi_{i}$ taking one of the four value $\left\{ -\frac{3\pi}{4},-\frac{\pi}{4},\frac{\pi}{4},\frac{3\pi}{4}\right\} $,
with $\left|A\right|$ minimizing 
\begin{equation}
\tilde{G}(\left|A\right|)=-\frac{H^{2}}{3}+2\left(\epsilon-\frac{{\epsilon H^{2}}}{6}+\frac{5a_{6}H^{4}}{216}\right)\left|A\right|^{2}+4\left(\frac{3\epsilon}{8}+\frac{25a_{6}H^{2}}{144}\right)\left|A\right|^{4}+\frac{5a_{6}}{3}\left|A\right|^{6}.\label{eq:G-tilde}
\end{equation}

The reduced potential $G$ is plotted in Figure \ref{fig:G-A-B} for
the case $\epsilon>0$. The structure has four non-trivial attractors due to a breaking of the symmetry imposed by the topography $h(y)=H\cos\left(2y\right)$.   In Figure \ref{fig:Minima-Saddle-StreamFunction}, we show the potential vorticity across $y$ for the two attractors, the corresponding saddle in between, and the topography.

\begin{figure}
\begin{center}
\includegraphics[width=0.6\columnwidth]{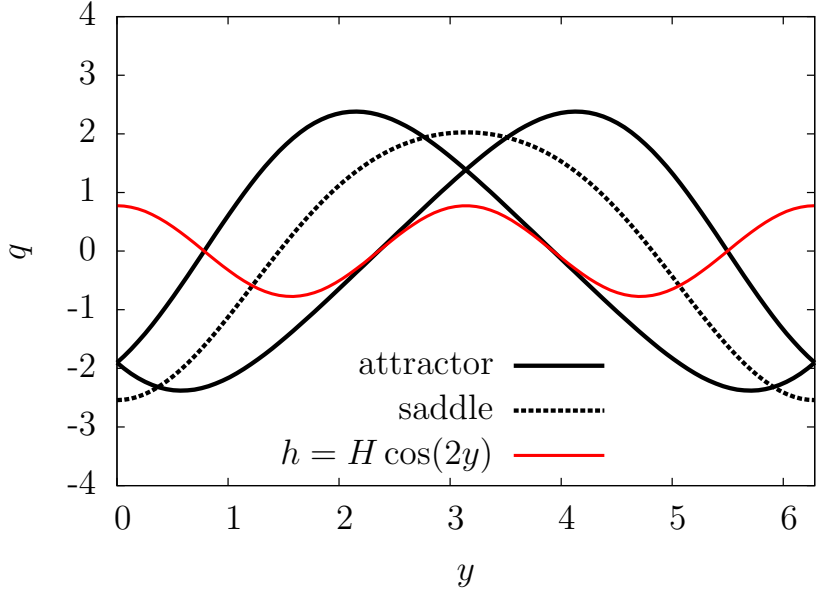} 
\caption{\label{fig:Minima-Saddle-StreamFunction}The plot depicts the topography
($h(y)=H\cos\left(2y\right)$, symmetric red curve) and two non-trivial attractors of the
potential vorticity $q$ (black solid lines) corresponding to two minima of the effective potential $G$ (see Eq. (\ref{eq:critical_points}),
and Figure \ref{fig:G-A-B}) for parameter value $\epsilon>0$. Additionally, we show the saddle between the two attractors
of the effect potential $G$ (dashed black curve). }
\end{center}
\end{figure}

Through the equilibrium hypothesis, we know how to describe and compute the the instantons corresponding to the phase
transitions between zonal flows. They are none other then the reversed trajectories for the relaxation paths for the dual dynamics.  The corresponding equation of motion for the relaxation paths for the dual
dynamics for the quasi-geostrophic dynamics has then been derived in section
\ref{sub:Instanton-equation-QG}. 

In the present example, the potential $\mathcal{G}$ is an even function, see Eq.
(\ref{eq:C-a4}). Also, we remark,
 that over the set of zonal flows $\mathbf{v}=U(y)\mathbf{e}_{x}$,
the nonlinear term of the quasi-geostrophic equation
vanishes: $\mathbf{v}\left[q+h\right]\cdot\nabla q=0$.  As a consequence,
when the instanton remains a zonal flow, the fact that $h$ has to
be replaced by $-h$ has no consequence and hence the dynamics will be time-reversible. Let us now argue that the
instanton is actually generically a zonal flow.

We assume for simplicity that the stochastic noise is homogeneous
(invariant by translation in both directions). Then $C\left(\mathbf{r},\mathbf{r'}\right)=C\left(\mathbf{r}-\mathbf{r'}\right)=C_{z}(y-y')+C_{m}(y-y',x-x')$
where 
\begin{equation}\label{eq:zonal_noise}
C_{z}(y)=\frac{1}{2\pi \delta}\int_{_{0}}^{2\pi \delta}\, C(x,y)\,{\rm d}x
\end{equation}
 is the zonal part of the correlation function, and $C_{m}=C-C_{z}$
the non-zonal or meridional part. 

As the nonlinear term of the two-dimensional Euler equations identically vanishes,
the relaxation dynamics has a solution among the set of zonal flows.
If $C_{z}$ is non-degenerate (positive definite as a correlation
function), then relaxation paths will exist through the gradient dynamics
\begin{equation}
\frac{\partial q}{\partial t}=-2\pi\alpha \delta\int_{_{0}}^{2\pi}\, C_{z}(y-y')\frac{\delta\mathcal{G}}{\delta q(y')}\, {\rm d}y',\label{eq:Relaxation-Path-Zonal}
\end{equation}
where $q=q(y)$ is the zonal potential vorticity field.

Moreover, as argued previously,
the fact that $G$ (\ref{eq:critical_points}) is a normal form for small enough, $a_{6}$, and $H$, implies that the gradient of $\mathcal{G}$
in directions transverse to $q=-A\cos\left(y\right)-B\cos\left(y\right)$ are much steeper
than the gradient of $G$. As a consequence, at leading order the
relaxation paths will be given by the relaxation paths for the effective
two-degrees of freedom $G$. Then, from (\ref{eq:Relation-G-GAB}),
(\ref{eq:critical_points}), and (\ref{eq:Relaxation-Path-Zonal})
we obtain that, at leading order, the dynamics of $A$ and $B$ are given by 
\begin{equation}
\frac{{\rm d}A}{{\rm d}t}=-c\frac{\partial G}{\partial A}\quad {\rm and}\quad \frac{{\rm d}B}{{\rm d}t}=-c\frac{\partial G}{\partial B},
\end{equation}
with $c=-\alpha \delta\int_{0}^{2\pi}\, C_{z}(y)\cos\left(y\right)\, {\rm d} y$, where we recall that $G$ is given by Eq. (\ref{eq:critical_points}).

From this result the relaxation paths are easily computed. Using the fact that
fluctuation paths are time reversed trajectories of relaxation paths, instanton
are also easily obtained. One of the resulting relaxation paths (blue curve) and
one of the instantons (red curve) are depicted in Figure \ref{fig:G-A-B}
overlapped on the contours of the potential $G$ in the $(A,B)$-plane. The corresponding two attractor involved, together
with the saddle point and examples of two intermediate states are shown in Figure \ref{fig:Instanton-IntermediateStates}.

\begin{figure}
\begin{center}
\includegraphics[width=0.6\columnwidth]{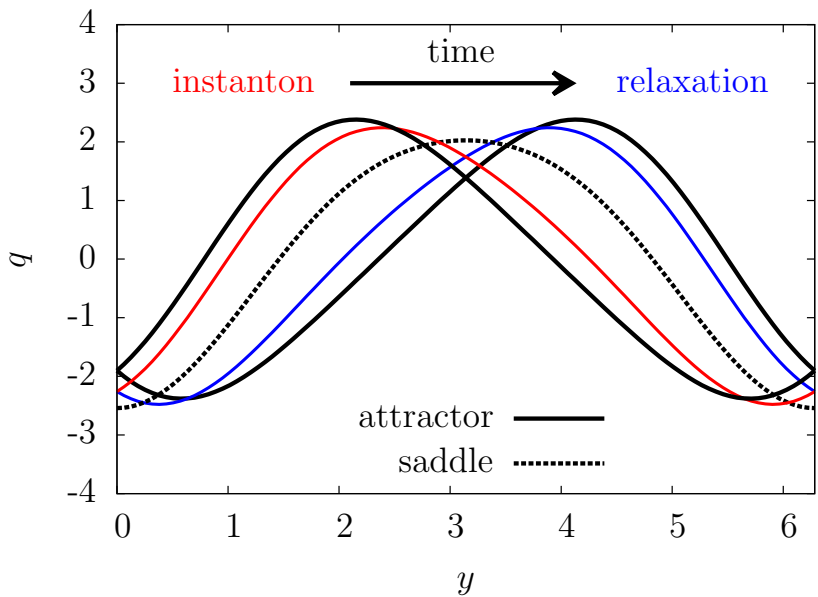}
\caption{\label{fig:Instanton-IntermediateStates} Figure taken from~\cite{bouchet_langevin_2014}. The potential vorticity
$q(y)$ for two of the non-trivial attractors (solid black curves), the corresponding saddle between the attractors (dashed black curve), and two intermediate
profiles along the instanton path (solid red curve) and the relaxation
path (solid blue curve). }
\end{center}
\end{figure}

For the Langevin dynamics formalism, the stationary probability distribution is known {\it a priori} and is given by
\begin{equation}
P_s[q] = \frac{1}{Z}\exp\left( -\frac{\mathcal{G}[q]}{\gamma} \right),
\end{equation}
where $Z$ is a normalization constant.  At a formal level, this can easily be computed by writing the Fokker-Planck equation for the evolution of the probability density functional.  Then the property that $P_s$ is stationary readily follows from the Liouville theorem and the fact that $\mathcal{G}$ consists of the conserved quantities of the deterministic dynamics.  

Subsequently, the transition rate $k$ for rare transitions between two attractors are going to be given by an Arrhenius law of the form
\begin{equation}\label{eq:arrhenius}
k = C\exp\left( \frac{\Delta\mathcal{G}[q]}{\gamma}\right),
\end{equation}
where $C$ is an order one prefactor and $\Delta\mathcal{G}=\mathcal{G}[q_{\rm saddle}]-\mathcal{G}[q_{\rm attractor}]$ is the potential difference between the saddle and the initial attractor. 

\subsection{Direct numerical simulations of rare transitions between coexisting attractors}
\label{sub:dns}

To verify the theoretical predictions of rare transitions in the equilibrium case, we perform direct numerical simulation of the Langevin example discuss above.

We numerically solve the Langevin dynamics of the quasi-geostrophic equations given by~\eqref{eq:qg_langevin} using a pseudo-spectral spatial discretization scheme of resolution $64\times 128$ on a periodic domain $[0,2\pi\delta)\times[0,2\pi)$ with $\delta = 1/2$.  Due to aliasing errors from the quintic nonlinearity associated to potential $\mathcal{G}$, we fully de-alaising courtesy of an 2/6th rule~\cite{gottlieb_numerical_1977}.  We time integrate the system using a second order Runge-Kutta method with time step $dt=2\times 10^{-3}$.  For simplicity we choose a white in time noise with correlation $C(\bx,\bxp)=\delta(\bx-\bxp)/Z$, where $Z$ is the normalization constant defined through condition~\eqref{eq:normalization}.   The use of such a noise correlation results in an ultraviolet divergence at high Fourier harmonics which may result in numerical instabilities. Therefore, to avoid a possible enstrophy bottleneck we include hyper-viscous dissipation to the right-hand side of Eq.~\eqref{eq:qg_langevin}, see Eq.~\eqref{eq:qg_langevin2}.  The addition of this extra dissipation breaks the equilibrium hypothesis on a general basis.  However, the dissipation only acts of the extreme high harmonics with little effect to the dynamics of the largest scales.  Therefore, we expect little deviation to the theoretical (equilibrium) prediction.  The numerical equation of motion is
\begin{subequations}\label{eq:qg_langevin2}
\begin{align}
\frac{\partial q}{\partial t}+\mathbf{v}\left[q-h\right]\cdot\mathbf{\nabla}q & =  -\alpha\int_{\mathcal{D}}\, C({\bf r},{\bf r}')\frac{\delta\mathcal{G}}{\delta q({\bf r}')}\,d{\bf r}' +\nu (-\Delta)^n(q-h)+\sqrt{2\alpha\gamma}\eta,\\
\mathbf{v}=&\mathbf{e}_{z}\times\mathbf{\nabla}\psi,\quad \omega=\Delta\psi, \quad q=\omega+h(\mathbf{r}),
\end{align}
\end{subequations}
As an additional check to the predictions of subsection~\ref{sub:Phase-transition-QG} we perform a relaxation of the system~\eqref{eq:qg_langevin2} from arbitrary initial conditions with $\nu=\gamma=0$.  From the effective potential landscape, the system should converge to the attractors of $\mathcal{G}$ given by~\eqref{eq:attractors}. By starting at four different regions of phase space, we indeed find the predicted attractors of $\mathcal{G}$ plotted in Figure~\ref{fig:numerical_attractors}. We observe that the theoretically predicted attractors (black dashed curves) overlay perfectly to the numerically founds attractors (coloured curves).

\begin{figure}[ht!]
\begin{center}
\includegraphics[width=0.6\columnwidth]{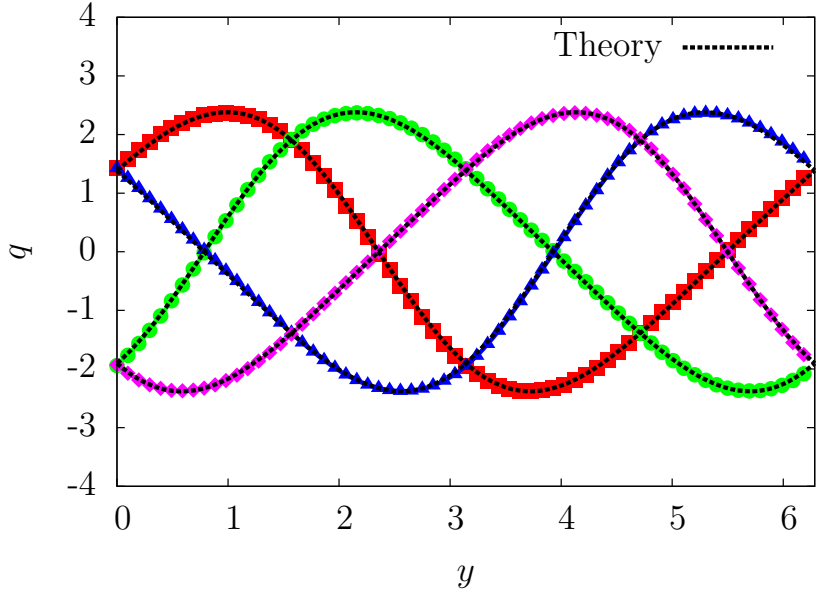}
\caption{\label{fig:numerical_attractors} Numerical attractors found using the relaxation equation~\eqref{eq:relaxation-paths-QG}.  The black dashed lines corresponds to the global minima from the effect potential $G(A,B)$. The red, blue, green and orange curves correspond to numerical solutions from Eq.~\eqref{eq:relaxation-paths-QG}.}
\end{center}
\end{figure}

\begin{figure}[ht!]
\begin{center}
\includegraphics[width=0.6\columnwidth]{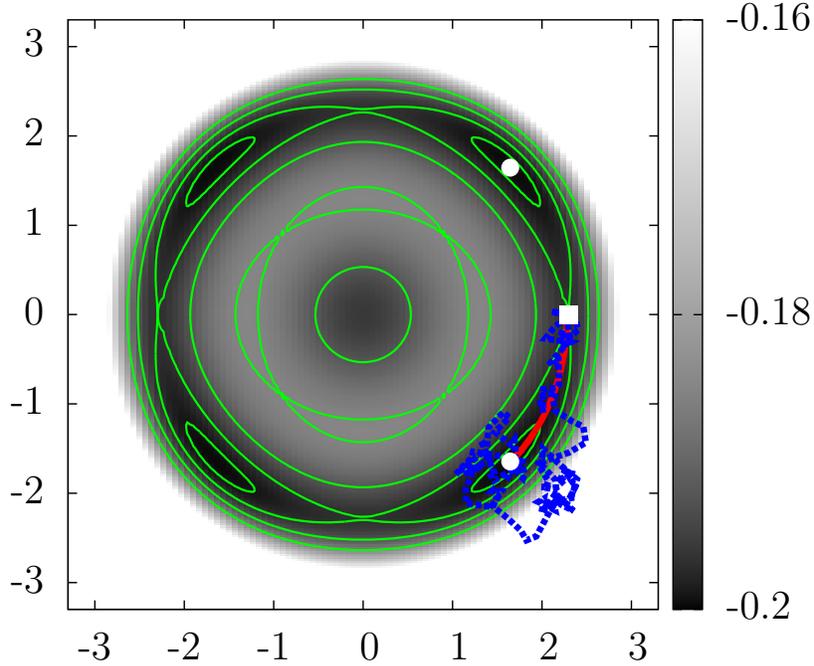}
\caption{\label{fig:dns_transition} Rare transition from one attractor to the neighbouring saddle with $\gamma=5\times 10^{-2}$ (blue dashed curve) taken from direct numerical simulation of system~\eqref{eq:qg_langevin2} overlaid over the contour plot of the effective potential landscape of Figure~\ref{fig:G-A-B} with the theoretically predicted instanton (solid red curve).}
\end{center}
\end{figure}

From the numerical perspective, we initialize the system beginning from one of the attractors and time step the system until we observe a transition to a neighbouring saddle.  We expect that if we are in a sufficiently weak noise limit $\gamma \ll\Delta \mathcal{G}$, then the transition in the direct numerical simulation will remain close to the theoretically predicted instanton (red curve in Figure~\ref{fig:G-A-B}).  Therefore, we utilize the following parameters in our numerical simulation: $\alpha = 1\times 10^{-1} $, $\nu = 1\times 10^{-13}$ and $\gamma = 5\times 10^{-2}$.

In Figure~\ref{fig:dns_transition} we plot the numerically observed transition onto the contour plot of the effective potential $G$.   We observe a relatively noisy transition (dashed blue curve) up to the saddle from the initial attractor.  We see a lot of the fluctuations at the base of the potential well were the gradients are small.  As the transition progresses up the potential well towards the saddle we observe better agreement to the theoretically predicted transition (solid red curve). We expect that much better agreement would be seen if we chose a smaller $\gamma$ however with the penalty of a far rarer transition.

\section{Numerical optimization of the Onsager-Machlup action functional for the barotropic quasi-geostrophic equations}
\label{sec:algorithm}

In this section we develop a numerical algorithm that computes local action minimizers of the Onsager-Machlup action functional defined in section~\ref{sec:models}.  These local action minimizers are candidates for the most probable transition paths between two states.  Unfortunately, these numerical optimization techniques are usually unable to distinguish between local minimizers and global ones. Therefore, using numerical schemes that are based on minimization of the action functional may not lead to the most likely transition path.  However, one may devise strategies to check to a certain degree whether a local or global minimum is obtained by perturbing found minima to see if they minimize to an alternative path.  If multiple minima exist, then comparison of the total action can be made.  As already stated, the numerical prediction of rare events without brute force simulations, observation or experiments is important.  The advantages of the action minimization methods are in that they can be applied to non-gradient systems in both equilibrium and non-equilibrium cases. 

Alternative strategies exist to compute rare transitions, such as string methods \cite{e_string_2002,e_simplified_2007},  the nudged elastic band method \cite{jonsson_nudged_1998}, eigenvector-following-type methods \cite{cerjan_finding_1981}, the dimer method \cite{henkelman_dimer_1999} and obtaining direct solutions of the instanton equations~\cite{grafke_instanton_2013}. However, many of these methods can not be applied to turbulent systems in general.

The numerical scheme that we implement here is based on the  adaption the {\it minimum action method} to turbulent systems.  In essence, the procedure uses a series of iterative estimates of transition paths until it finds a local minimum of the action functional $\mathcal{A}$.  This numerical method is applicable to both the equilibrium and non-equilibrium cases and can easily be extended to consider more complex turbulence problems.

\subsection{The minimum action method}
\label{sub:mam}

The minimum action method, is a class of numerical optimization procedures that determine local minimizers of functionals.  One of the key properties of the algorithm is that it can be applied to systems that do not provide {\it a priori} an energy potential landscape.  This makes it ideal to study rare transitions in turbulence problems. It has already found many applications in the use of computing most probable transition paths in low dimensional gradient systems \cite{e_minimum_2004,zhou_adaptive_2008}, rare transitions in the the Kuramoto-Siavashinksky equation~\cite{wan_study_2010} and the Kardar-Parisi-Zhang equation~\cite{fogedby_minimum_2009}. In this section, we will outline an algorithm for the standard minimum action method, however there exists many more advanced versions they may be useful in the future.  These include algorithms that provide adaptive re-meshing known as {\it adaptive minimum action methods}~\cite{zhou_adaptive_2008,wan_adaptive_2011} or one that use an arc length parameterization of time in order to compute infinite time transition paths, known as {\it geometric minimum action methods}~\cite{heymann_geometric_2008,vanden-eijnden_geometric_2008}. 

The generic strategy of the minimum action method algorithm is to begin with an initial estimate of the most probable transition path between two states.  Then, with the use of variations of the action functional with respect to the transition trajectory, improvements in the form of iterations to the initial guess can me made that subsequently reduces the action.  This iterative process is continually repeated until the series of estimates converges to a local minima of the action functional.

The main complexity of this method is in determining how one should improve each guess so that the action is reduced.  There are various strategies one may use, such as applying Newton's method~\cite{nocedal_numerical_2006}, which uses information about the first and second variations (Hessian) of the action functional or quasi-Newton methods, such as the popular Broyden-Fletcher-Goldfarb-Sahnno (BFGS) scheme that iteratively approximates the Hessian without the need to compute it directly. Newton's method is a relatively expensive procedure especially for high dimensional minimization problems where the computation of the Hessian is difficult. Subsequently, quasi-Newton methods have been favoured in the community and has been successfully applied to the minimum action method but for only in situations of low dimensional gradient systems~\cite{e_minimum_2004,fogedby_minimum_2009}.  

We found that for turbulence problems, where we have to deal with a large number degrees of freedom, even quasi-Newton methods are expensive.  Therefore, we are resorted to relying on methods based solely using the first variation of the action functional.  The simplest method that falls into this category is the method of steepest descent, where a descent direction $d$ is taken in the direction of the local anti-gradient of the action functional, i.e. $d=-\delta\mathcal{A}/\delta q$.  Usually, these methods can have poor convergence rates when the potential energy landscape consists of long narrow valleys, where the minimization procedure leads to zig-zagging across the narrow valley rather than along it.   To improve convergence in these situations, one can use the nonlinear conjugate gradient method, which uses knowledge about previous descent steps to avoid crossing back and forth across potential valleys~\cite{nocedal_numerical_2006}.  It is with this in mind that we utilize a nonlinear conjugate gradient method for our problem.

For out notation, we label each iteration with a superscript such that the $n^{\mathrm{th}}$ estimate of the most probable transition path is labeled as ${q}^{n}$ for $n=0,1,2,\dots$.  The initial guess is denoted as $q^0$.   Each new estimate for the most probable transition path ${q}^{n+1}$ is computed from the previous guess $q^{n}$ by taking an appropriate descent step of size $l^n$ in the descent direction $d^n$:
\begin{equation}\label{eq:iteration}
q^{n+1} = q^n + l^n d^n.
\end{equation}
The descent direction $d^n$ is obtained through a nonlinear conjugate gradient method~\cite{nocedal_numerical_2006}. In general, for nonlinear conjugate gradient methods one takes the descent direction as
\begin{equation}\label{eq:descent}
d^{n+1} = -\frac{\delta \mathcal{A}}{\delta {q}^n} + \beta^n d^n \quad {\rm where} \quad d^0=-\frac{\delta \mathcal{A}}{\delta {q}^0}.
\end{equation}
The parameter $\beta^n$ is known as the nonlinear conjugate gradient parameter which determines to what level the current descent direction should depend on the previous descent direction.  There are various ways of computing $\beta^n$, but we found that the most optimal was to use the standard Fletcher and Reeves formula where
\begin{equation}
\beta^n = \frac{ \left|\left| \frac{\delta \mathcal{A}}{\delta q^n} \right|\right|^2}{\left|\left|\frac{\delta \mathcal{A}}{~\delta q^{n-1}} \right|\right|^2},
\end{equation}
where $||\cdot||$ is an appropriate norm. Due to the finite number of degrees of freedom associated to the numerical discretization, there can only be a finite number of orthogonal descent directions.  Therefore, it will become important to occasionally reset the nonlinear conjugate gradient parameter when to consecutive descent directions are far from being orthogonal.  To achieve this, we set $\beta^n=0$, resulting in standard steepest descent step when
\begin{equation}
\frac{ \left( \frac{\delta \mathcal{A}}{\delta q^n}, \frac{\delta \mathcal{A}}{~\delta q^{n-1} }\right)}{\left|\left|\frac{\delta \mathcal{A}}{~\delta q^{n-1}} \right|\right|^2} > 0.5,
\end{equation}
where $\left(\cdot,\cdot\right)$ is the inner product associated to the norm $||\cdot||$.

The step length $l^n$ is chosen such that we obtain the the greatest reduction in the action functional.  In order to ensure that $d^n$ corresponds to a descent direction (that it results in the reduction of the action), the step length $l^n$ must satisfy the strong Wolfe conditions~\cite{nocedal_numerical_2006}. Fortunately, there exists standard line search algorithms in order to determine the largest step length that satisfies the strong Wolfe conditions.  Therefore, we implement the line search algorithm 3.5 of chapter 3 in~\cite{nocedal_numerical_2006}.

The minimization is continuously performed for each iteration until the estimate of the most probable transition trajectory $q^n$ is within some tolerance, say $\epsilon$ of being a solution of the Euler-Lagrange equations~\eqref{eq:Euler-Lagrange}.  This is checked by halting the algorithm if the solution satisfies the condition
\begin{equation}
	\left|\left| \frac{\delta \mathcal{A}}{\delta q^n} \right|\right| < \epsilon.
\end{equation}

\subsection{Numerical discretization of the action functional}

In order the numerically minimize the action functional, we must first discretize the action in both time and space.  Due to the periodicity of our domain, it is natural for us to consider the Fourier harmonics as the standard basis.  In time, we approximate the transition on a uniform grid of $N_t +1$ points along the interval $[0,T]$. All spatial derivatives are computed in Fourier space with the nonlinear terms computed in physical space using the 2/3rds dealiasing rule (the standard pseudo-spectral method~\cite{gottlieb_numerical_1977}).  Derivatives in time are achieved by applying the second-order central finite difference scheme upon a staggered grid labelled by $\{j+1/2\}$ for $j=0,\dots, N_t-1$, where time is parameterized by $t_j = j\Delta t$ for $j=0,\dots, N_t$ and $\Delta t = T/N_t$.  

In this respect, the transition trajectory is fully represented by the set of Fourier amplitudes $\left\{q_{\bk,j}\right\} \equiv \left\{q_{\bk}(j\Delta t)\right\}$ given by Eq.~\eqref{eq:basis}, for $j=0,\dots, N_t$ and $\bk = \left(2\pi n_x/L_x,2\pi n_y/L_y\right)$ for $n_{x,y} \in \left\{-N_{x,y}/2,\dots,N_{x,y}/2 -1\right\}$ where $N_x\times N_y$ is the spatial resolution. Due to the reality of the potential vorticity $q$, the Fourier harmonics satisfy the condition $q_\bk = q_{-\bk}^*$ at every point in time.

Using this convention, we define the numerically discretized action functional $A[q]$ as
\begin{equation}\label{eq:ActionD}
A[q] =  \frac{\Delta t}{2} \sum_{j=0}^{N_t} \sum_{\bk}  \frac{\left|\dot{q}_{\bk,j+\frac12} +  \left(\bv \cdot \bn q\right)_{\bk,j+\frac12} +\alpha\omega_{\bk,j+\frac12} +\nu k^{2n} \omega_{\bk,j+\frac12}\right|^2}{|f_\bk|^2},
\end{equation}
where we have used the notation $\dot{q}_{\bk,j+\frac12} \equiv \left(q_{\bk,j+1}-q_{\bk,j}\right)/\Delta t$ to denote the time derivative defined on the staggered grid.  As a consequence, we must also compute the linear and nonlinear terms on the same grid.  To do this, we average the contribution of neighbouring points by simple interpolation $q_{\bk,j+1/2} = \left(q_{\bk,j+1}+q_{\bk,j}\right)/2 $.  

To compute the first variation of the action functional $\delta \mathcal{A}/\delta q$, we express the action in terms of its Lagrangian:
\begin{equation}
\frac{\delta {A}}{\delta {q}_{\bk,j}} =   \frac{\Delta t}{2}\left(\frac{\delta  {L}}{\delta {q}_{\bk,j+\frac12}}+  \frac{\delta  {L}}{\delta {q}_{\bk,j-\frac12}}\right) -  \left( \frac{\delta {L}}{\delta \dot{{q}}_{\bk,j+\frac12}}  -  \frac{\delta  {L}}{\partial \dot{{q}}_{\bk,j-\frac12}}\right), 
\end{equation}
where the variations of the Lagrangian are explicitly given as 
\begin{equation}
\frac{\delta {L}}{\delta \dot{{q}}_{\bk,j+\frac12}}\equiv \; p_{\bk,{j+\frac{1}{2}}} = \frac{1}{|f_\bk|^2}\left[\dot{{q}}_{\bk,j+\frac12} + {(\bv \cdot \bn q)}_{\bk,j+\frac12}  +\alpha{\omega}_{\bk,j+\frac{1}{2}}+ \nu k^2{\omega}_{\bk,j+\frac{1}{2}} \right]
\end{equation}
and
\begin{equation}
\frac{\delta {L}}{\delta {q}_{\bk,j+\frac{1}{2}}} =  -{\left(\bv \cdot \bn p\right)}_{\bk,j + \frac{1}{2}} +{\left\{\Delta^{-1}\left[ \left(\bn q_{j+\frac{1}{2}} \times \bn p_{j + \frac{1}{2}}\right)\cdot \be_z \right]\right\}}_{\bk,j+\frac12}+\alpha {p}_{\bk,j+\frac{1}{2}} + \nu k^2 {p}_{\bk,j+\frac{1}{2}}.
\end{equation}
Notice that the expressions $p$, $\left(\bv\cdot\bn p\right)$ and $\Delta^{-1}\left[\left(\bn q\times\bn p\right)\cdot\be_z\right]$, are only defined on the staggered grid indexed by $\{j+{1/2}\}$.  Consequently, our notation is defined as $\left(\bv \cdot \bn p\right)_{\bk,j + \frac{1}{2}} = \left(\bv_{j+\frac12} \cdot \bn p_{j+ \frac12}\right)_{\bk}$. To evaluate these quantities back onto the original grid we interpolate the quantities by $p_{\bk,j} = (p_{\bk,j+\frac{1}{2}} + p_{\bk,j-\frac{1}{2}})/2$.

Finally, after some straighforward mathematics, we arrived and the numerical expression for the first variation of the action function with respect to $q$:
\begin{equation}\label{eq:action_derivative}
\frac{\delta {A}}{\delta {q}_{\bk,j}} = \Delta t \left[ \dot{{p}}_{\bk,j} - {\left(\bv\cdot\bn p\right)}_{\bk,j} +{\left\{\Delta^{-1}\left[\left(\bn q\times\bn p\right) \cdot \be_z\right]\right\}}_{\bk,j} + \alpha p_{\bk,j} + \nu k^2 p_{\bk,j}\right]
\end{equation}
where the time derivative of $p$ is the standard central finite difference expression $\dot{{p}}_{\bk,j}=(p_{\bk,j-\frac{1}{2}} - p_{\bk,j+\frac{1}{2}})/ {\Delta t}$.

\section{Numerical predictions for the most probable rare transitions}
\label{sec:numerical_trans}

To show that the minimum action method is suitable for the prediction of rare transitions in turbulent models, we consider a series of examples that verifies the algorithm.  In this section we will begin by considering the over-damped limit of the barotropic quasi-geostrophic dynamics where the nonlinearity is assumed to be absent.  We follow with an example that satisfies the equilibrium hypothesis of section~\ref{sec:equilibrium} in a regime of a single global dynamical attractor.  Through this example, we check the numerically obtained transition agrees with the prediction made through the equilibrium theory. Finally, we consider a geophysical based example of considering a rare transition between two distinct zonal jet configurations modelled by the  quasi-geostrophic equations.  In all cases, we show good agreement of the numerical prediction to analytical predictions.

\subsection{The over-damped limit}
\label{sub:overdamp}

The over-damped limit $\alpha, \nu\gg 1$ of the barotropic quasi-geostrophic equations corresponds to dynamics that are dominated by dissipative effects. Therefore, we can make the assumption that the nonlinearity is sub-dominant and can be neglected.  Moreover, for simplicity, we assume also that the topography is absent $h=0$. We remark that this limit, although unphysical in reality provides a simple way of verifying the minimization procedure of the minimum action method.  Absence of the nonlinearity reduces the system to a linear problem allowing for a theoretical treatment. Indeed, the instanton equations~\eqref{eq:Euler-Lagrange} become a series of linearly independent differential equations for each Fourier amplitude that can be straightforwardly solved.  The over-damped dynamics are given as
\begin{subequations}
\begin{align}\label{eq:over-damped_dynamics}
\frac{\partial \omega}{\partial t}  =& -\alpha \omega - \nu \left(-\Delta\right)^n \omega + \sqrt{2\alpha\gamma} \eta,\\
\mathbf{v}=&\mathbf{e}_{z}\times\mathbf{\nabla}\psi,	\qquad \omega=\Delta\psi,
\end{align}
\end{subequations}
Due to the linearity of Eq.~\eqref{eq:over-damped_dynamics}, the dynamics of the Fourier representation of the vorticity field $\omega$ can be represented as a series of uncoupled Ornstein-Uhlenbeck processes for each Fourier amplitude $\omega_{\bk}$.  This linearity means that the instanton equations can be directly solved yielding a theoretical prediction for a rare transition between two states for a transition time $T$.  By working in the Fourier representation, the instanton equations can be reduced to series of second-order, linear boundary value problem for each Fourier amplitude
\begin{subequations}\label{eq:EL-overdamp}
\begin{align}
\frac{\partial^2 {\omega}_\bk}{\partial t^2} = &\left(\alpha + \nu k^2\right)^n{\omega}_\bk,\\
{\omega}(\bk,0) = {\omega}_\bk(0), &\quad {\omega}(\bk,T) = {\omega}_\bk(T).
\end{align}
\end{subequations}
Notice that the noise correlation $|f_{\bk}|$ drops out of the instanton equations, meaning that the trajectory is independent on the noise--this is a consequence of the decoupling of each Fourier amplitude from each other.  Eqs.~\eqref{eq:EL-overdamp} can be readily solved with solution
\begin{subequations}\label{eq:overdamped_sol}
\begin{equation}
\omega^*(\bx,t) = \sum_{\bk} \, \omega_{\bk}^*(t)\,\be_{\bk}(\bx),
\end{equation}
with
\begin{equation}
{\omega}_\bk^*(t) = \frac{\sinh(\beta_\bk[T-t]){\omega}_\bk(0) + \sinh(\beta_\bk t){\omega}_\bk(T)}{\sinh(\beta_\bk T)},
\end{equation}
\end{subequations}
where for we have used the notation $\beta_\bk = \alpha+\nu k^{2n}$.  As one can observe from solution~\eqref{eq:overdamped_sol}, the most probable rare transition corresponds to an exponential decay from the initial state with rate $\beta_{\bk}$ followed by an exponential increase with the same rate to the final state.  In essence, the transition wants to decay to the zero state with the relaxation defined by the dissipation rate $\beta_{\bk}$ for each Fourier mode.  Once decayed, the trajectory will transition to the final state with an exponential rate governed again by the dissipation rate.  It should be noticed, that for large transition times $T$, the majority of the transition will result in the state being close to zero, with most of the dynamics occurring at the beginning and at the end of the transition on a timescale defined by the dissipation rate $\beta_{\bk}$

The corresponding Lagrangian for the theoretically predicted rare transition~\eqref{eq:overdamped_sol} is given by
\begin{align}\label{eq:lag-overdamp}
\mathcal{L}\left[\omega^*,\frac{\partial \omega^*}{\partial t}\right] =& \frac{1}{2} \int_{\mathcal{D}} \int_{\mathcal{D}}\left[\frac{\partial \omega^*}{\partial t} + \alpha \omega^* + \nu\left(-\Delta\right)^n \omega^* \right](\bx)\nonumber
\\&\times C^{-1}(\bx,\bx') \left[\frac{\partial \omega^*}{\partial t} + \alpha \omega^* + \nu\left(-\Delta\right)^n \omega^* \right](\bx') \, d\bx \, d\bx'\nonumber\\
=&\frac{1}{2}\sum_{\bk} \frac{\beta_\bk \exp\left(2\beta_{\bk} t\right)}{|f_\bk|^2\sinh^2\left(\beta_{\bk}T \right)}\left| \omega_{\bk}(T)-\omega_{\bk}(0)\exp\left(\beta_{\bk} T\right)\right|^2.
\end{align}

The Lagrangian~\eqref{eq:lag-overdamp} quantifies how much momentum is required from the noise to push the transition to the final state. Therefore, it is an important quantity for the rare transition characterizing the effect of the noise along the transition and also yields the action upon time integration. 

We now test the minimum action method and compare the numerically obtained transition path to that predicted by the theory.  We apply the numerical method to the over-damped system defined above. We select the initial and final states to be $\omega_0=[\cos(x)-(2/5)\sin(x)+(1/5)\cos(y)+(3/5)\cos(x+y)-(4/5)\sin(2y-x)]/E$ and $\omega_T=[(1/2)\cos(x)+(2/5)\sin(x)+(3/5)\cos(3y)-(1/5)\cos(2y-x)+(1/5)\sin(2y-x)]/E$ appropriately normalized through $E$ to give unit energy density.  The two states are displayed in Figure~\ref{fig:initial_final} and were chosen so that they contain a large number of modes.  We perform the minimization with an initial trajectory defined through linear interpolation between the two boundary states in time.  We use $N_x=N_y=16$ Fourier modes and a temporal grid of $N_t=101$ points for a periodic spatial domain of size $L_x=L_y=2\pi$ and time domain of length $T=10$. The dissipation parameters that we use are $\alpha = 1\times 10^{-1} $ and $\nu = 5\times 10^{-2}$ with $n=1$.  We choose a noise correlation that represents a Gaussian white noise with $C(\bx-\bxp)=\delta(\bx-\bxp)/Z$ where $Z$ is the normalization constant to ensure relation~\eqref{eq:normalization} holds.

Displayed in Figure~\ref{fig:w_hat} (left) is the time evolution of absolute value of each Fourier mode in the transition, and (right) the complex phase space of the transition of each Fourier mode.  In both plots, the theoretical predictions arising from Eq.~\eqref{eq:overdamped_sol} are overlaid by the black dashed curves.  We observe excellent agreement between the numerical and theoretical results.  We observe a slight discrepancy to the numerical data in the time evolution of mode $\bk =(-1,2)$, however this is certainly down to numerical resolution close to the cusp where the transition goes from exponential decay to growth near $t=5$.  In Figure~\ref{fig:w_hat} (right), we quite clearly observe that the transition quickly decays to the zero state for each Fourier amplitude before transitioning to the final state. In Figure~\ref{fig:lag_t} we plot the time evolution of the Lagrangian for the numerical prediction and compare to the theoretical result given by~\eqref{eq:lag-overdamp}.  We observe excellent agreement to the theory and see that the majority of the Lagrangian is appears at later times where the transition needs to be pushed against the dissipation to reach the final state.

\begin{figure}[ht!]
\begin{center}
\hfill
\includegraphics[width = 0.4\columnwidth]{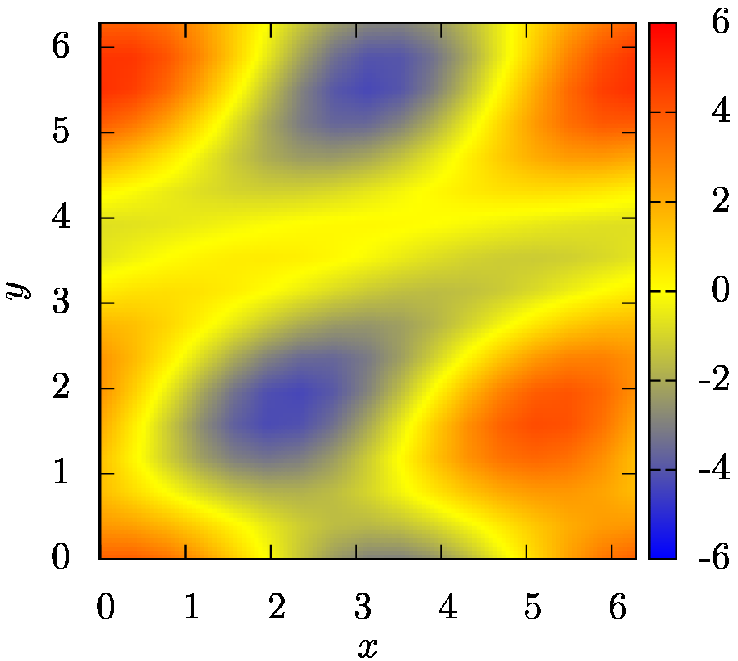}
\hfill
\includegraphics[width = 0.4\columnwidth]{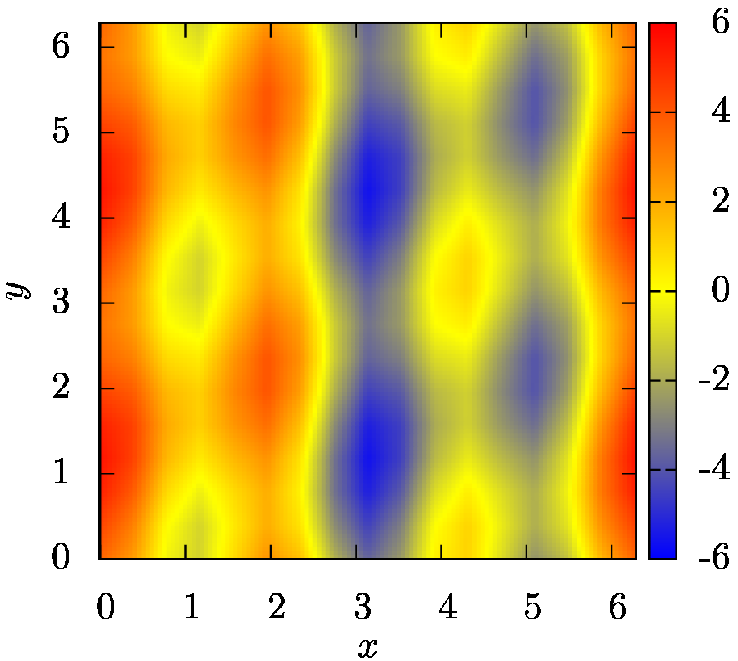}
\hfill
\caption{We plot the vorticity distribution of the initial $\omega_0=[\cos(x)-(2/5)\sin(x)+(1/5)\cos(y)+(3/5)\cos(x+y)-(4/5)\sin(2y-x)]/E$ and final  $\omega_T=[(1/2)\cos(x)+(2/5)\sin(x)+(3/5)\cos(3y)-(1/5)\cos(2y-x)+(1/5)\sin(2y-x)]/E$ states appropriately normalized through the constant $E$ to give unit energy density.\label{fig:initial_final}}
\end{center}
\end{figure}

\begin{figure}[ht!]
\begin{center}
\hfill
\includegraphics[width = 0.45\columnwidth]{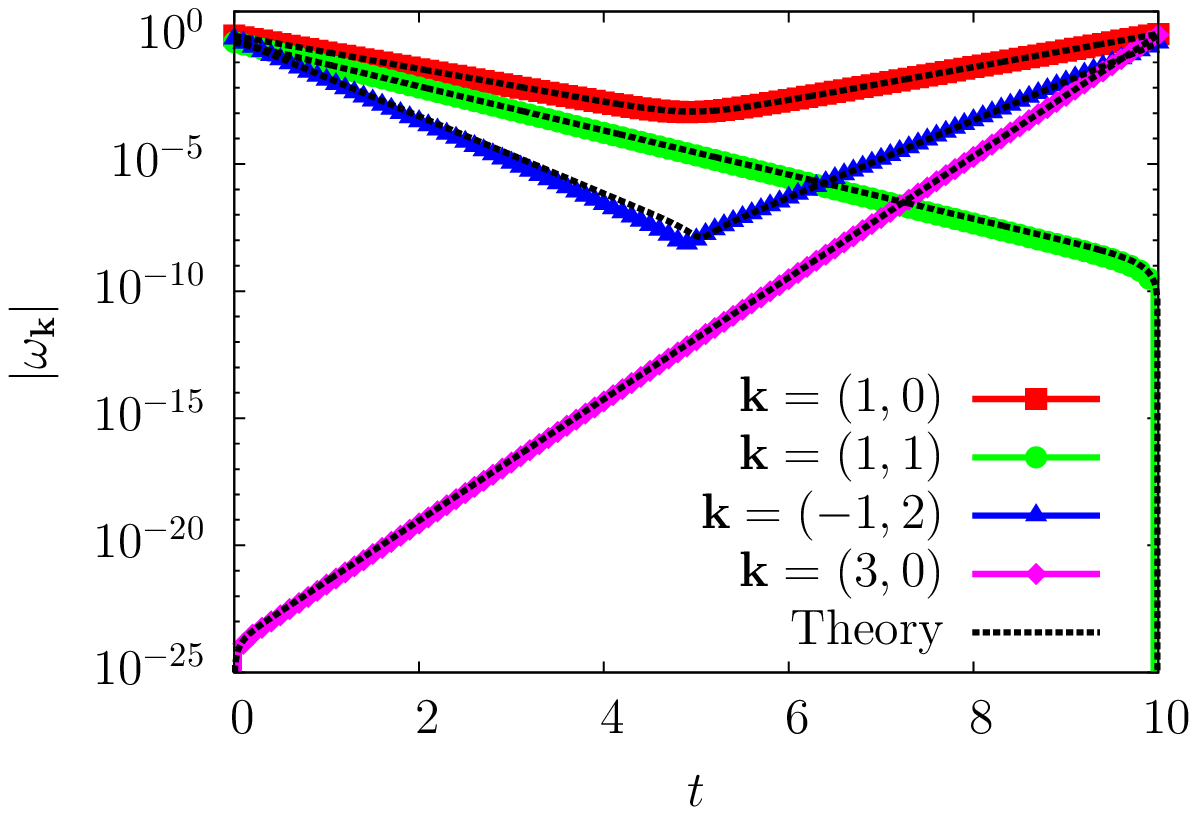}
\hfill
\includegraphics[width = 0.45\columnwidth]{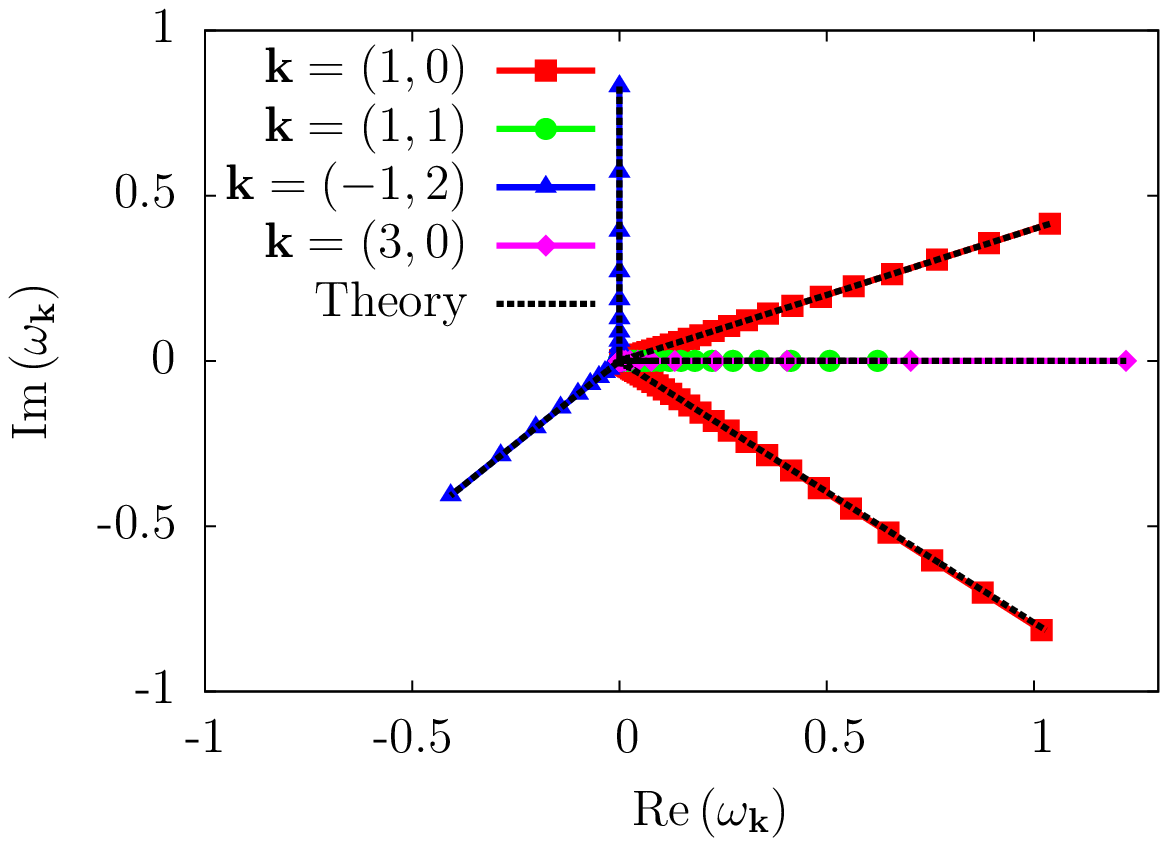}
\hfill
\caption{(Left) we plot the time evolution of $|\omega_{\bk}(t)|$ for each mode $\bk$ and (Right) we plot the complex phase space trajectories of each mode.  For each, the theoretical prediction of Eq.~\eqref{eq:overdamped_sol} are overlaid by the black dashed curves.\label{fig:w_hat}}
\end{center}
\end{figure}

\begin{figure}[ht!]
\begin{center}
\includegraphics[width = 0.5\columnwidth]{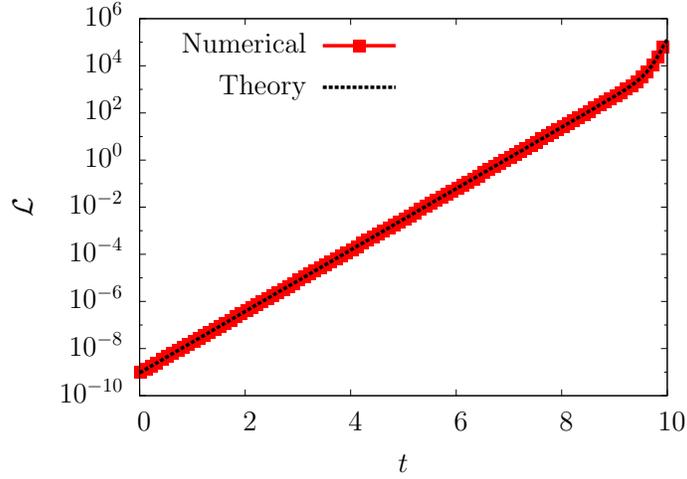}
\caption{Time evolution of the Lagrangian $\mathcal{L}$ for the over-damped problem.  The theoretical result of~\eqref{eq:lag-overdamp} is plotted by the dashed black curve.\label{fig:lag_t}}
\end{center}
\end{figure}

We conclude that the numerical minimization for the over-damped system yields the expected results predicted through the Freidlin-Wentzell theory. However, we stress that this example ignores the effect of the nonlinear advection term of the quasi-geostrophic equations, which we will discuss in the next subsections.

\subsection{Equilibrium instanton starting at zero}

In this subsection, we consider applying the minimum action method to an example that satisfies the equilibrium hypothesis of section~\ref{sec:equilibrium}. Such an examples will allow for the direct comparison to the predictions made in section~\ref{sec:equilibrium} verifying, not only the numerical optimization algorithm but also the equilibrium theory.  Our setup will be the following: we will consider a transition beginning at the zero state and transitioning to another, non-zero, state.  What is essential is that we compute this transition in the equilibrium regime where there is only one global dynamical attractor, the zero state.  This is important because we wish to compare the numerical prediction to the solution defined through the relaxation from the time-reversed transition in the corresponding dual dynamics defined through the relaxation equation~\eqref{eq:relaxation-paths-QG}.  The criterion of zero being the only attractor is important as this comparison can only be made if the transition remains in the same basin of attraction to that of the attractor in which the transition starts. (Transitions that occur across several basin of attractions will have to be compared to a theoretical transition composed of several instantons and relaxation trajectories corresponding to each attractor and saddle the transition passes through.)  By considering a setup with only one attractor, then all possible non-zero states will be within the basin of attraction of $\omega=0$ and the transition can be cmopared to one instanton prediction through the relaxation equations.

%We now consider the action minimization of the full nonlinear system in the equilibrium regime satisfying the condition of section~\ref{sec:equilibrium}. Here, we can consider two examples of equilibrium rare transitions.  The first, arises through a specific example where the the initial and final states are eigenfunctions of the Laplace operator with the same eigenvalue.  This assumption results in the problem becoming an equilibrium one resulting in the most probable rare transition being one defined through a continuous set of steady states composed of these eigenfunctions. The second example is more interesting, we consider the rare transition from zero to any non-zero state and compare the numerically obtained transition predicted through the minimum action method to the time-reversed relaxation path obtained through Eq.~\eqref{eq:relaxation-paths-QG}.  This is important for two reasons; one that in the limit of $T\to \infity$ the minimized trajectory should coincide with the trajectory obtained from the relaxation equation, confirming the equilibrium hypothesis, and two it verifies that the numerical code works in situations where nonlinearity is important.

 % % % % % % % % % % % % % % % % % %Second Example % % % % % % % % % % % % % % % % % % % % % % % % % %
In general determining a transition from zero to an arbitrary state $\omega_T$ analytically is difficult. However, by ensuring  the equilibrium hypothesis holds will helps us in this regard.  We know from section~\ref{sec:equilibrium} that the rare transition from an attractor to any state within the basin of attraction of that attractor will be the time-reversed relaxation path of the corresponding dual system. Therefore, by considering the relaxation path Eq.~\eqref{eq:relaxation-paths-QG}, we will recover the instanton: the most probable infinite time fluctuation path. Unfortunately, due to the numerical discretization of the minimum action method, we are unable to ascertain the infinite time transition path.  However, one would expect that if $T$ is sufficiently large enough then the two transitions should be relatively close.  Therefore, we will consider a sequence of transitions for increasing $T$ and show convergence to the instanton. 

The equilibrium setup is as follows: we consider a noise spectrum that is uniform in Fourier space, i.e. corresponding to a Gaussian white noise with a correlation $C(\bx,\bxp)=\delta(\bx-\bxp)/Z$ and a potential that is proportional to the enstrophy measure $\mathcal{G}\propto \omega^2/2$.  This is important for three reasons, {\it i)} this corresponds to linear friction in the two-dimensional Euler equations meaning that the model is realistic in some sense, {\it ii)} this potential and the noise correlation satisfy the equilibrium hypothesis of section~\ref{sec:equilibrium}, and {\it iii)} the quadratic form of the potential implies that there is only one minimum corresponding in this case to the zero state $\omega=0$.

For the numerics, we choose the final state to be $\omega_T = [\cos(x)-(2/5)\sin(x) + (2/5)\cos(x+y)-(1/2)\sin(x+y)\allowbreak+(3/5)\cos(2x+y)-(1/5)\sin(2x+y)]/E$, where $E$ is the normalization constant to give unit energy density. 

We use a Fourier resolution of $N_x=N_y=16$ Fourier harmonics in the minimum action method and compare a series of minimizations with increasing transition time $T$.  In each realization we ensure that we have sufficient temporal resolution by using a fixed grid spacing of $T/N_t= 10^{-1}$.

In Figure~\ref{fig:w_and_energy_decay} we show the vorticity distribution of the final transition state (left) and the time evolution of the transition energy $\mathcal{E}$ (right).  Notice, that from the energy balance equation~\eqref{eq:energybalance}, we expect an exponential decrease of the energy with rate $2\alpha$.  This is exactly what is observed from the relaxation trajectory (black dashed line in Figure~\ref{fig:w_and_energy_decay}).  Moreover, observe that the numerical minimization predictions also agree with this decay rate initially. The discrepancy at later times is a consequence of the minimization procedure only dealing with transition of finite transition time $T$, such that the energy must vanish in finite time.  This is also supported by the observation that for increasingly longer transition times results in better agreement to the expected energy decay.  Of course one expects, and is indicated by the numerics, that this agreement will be exact in the limit of $T\to \infty$.  

\begin{figure}[ht!]
\begin{center}
\hfill
\includegraphics[width = 0.35\columnwidth]{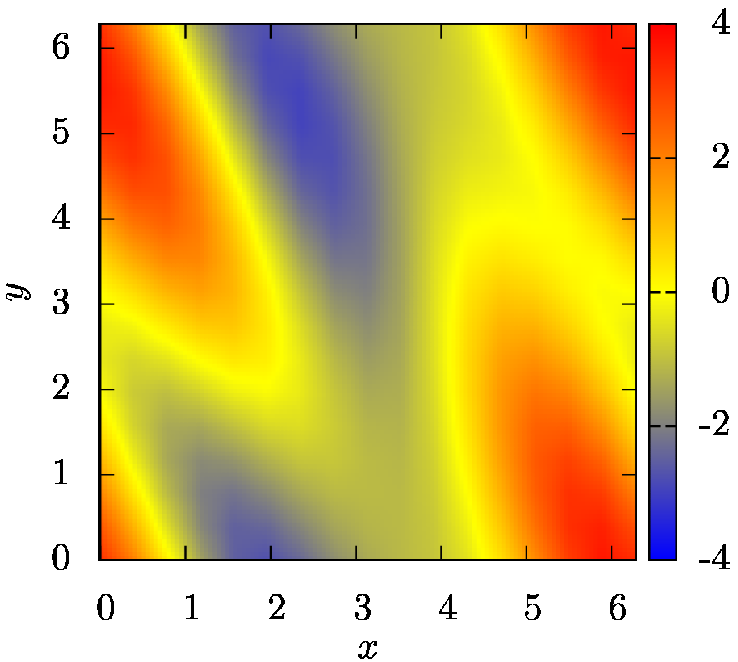}
\hfill
\includegraphics[width = 0.45\columnwidth]{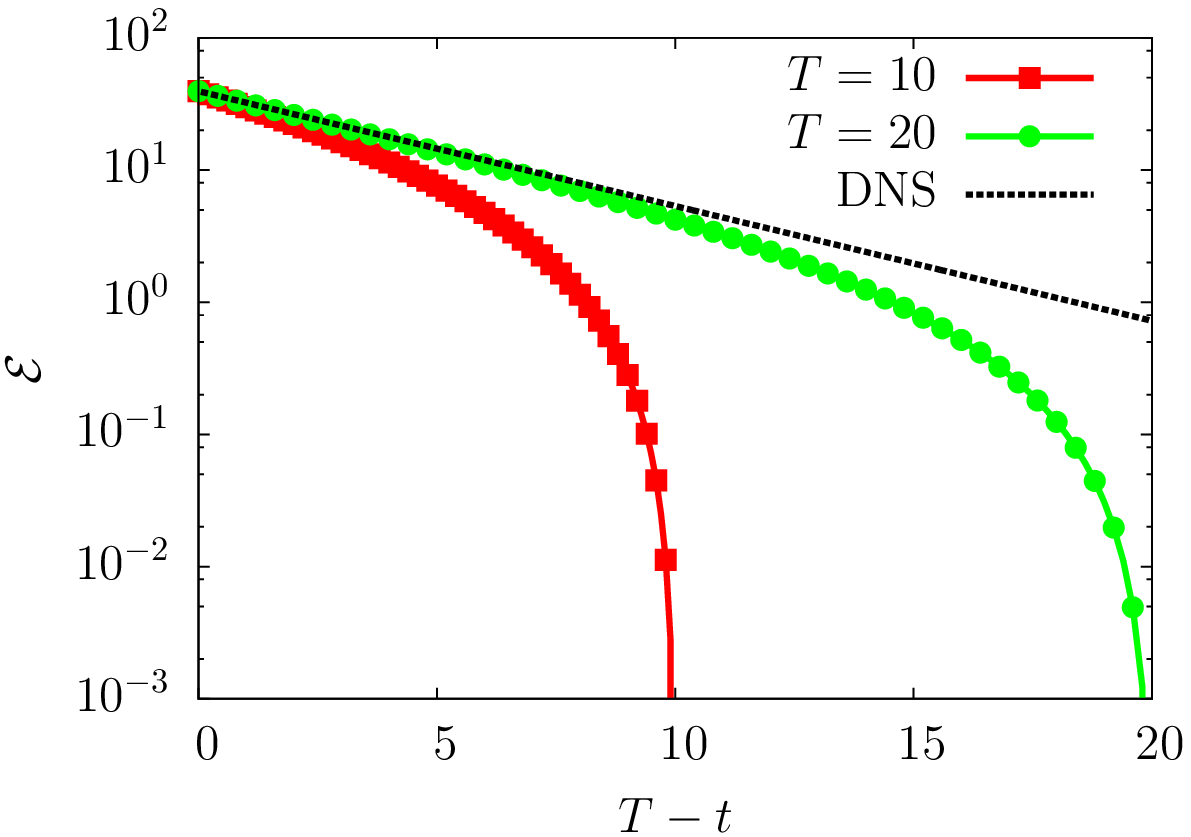}
\hfill
\caption{(Left) The final transition state $\omega_T = [\cos(x)-(2/5)\sin(x) + (2/5)\cos(x+y)-(1/2)\sin(x+y)+(3/5)\cos(2x+y)\allowbreak-(1/5)\sin(2x+y)]/E$ where $E$ is a normalization constant to give unit energy density.  (Right) The time evolution of energy of the predicted transition path for transition times $T=10$ and $T=20$ compared to the relaxation trajectory from direct numerical simulation corresponding to the $T\to\infty$ limit.\label{fig:w_and_energy_decay}}
\end{center}
\end{figure}

\begin{figure}[ht!]
\begin{center}
\hfill
\includegraphics[width = 0.3\columnwidth]{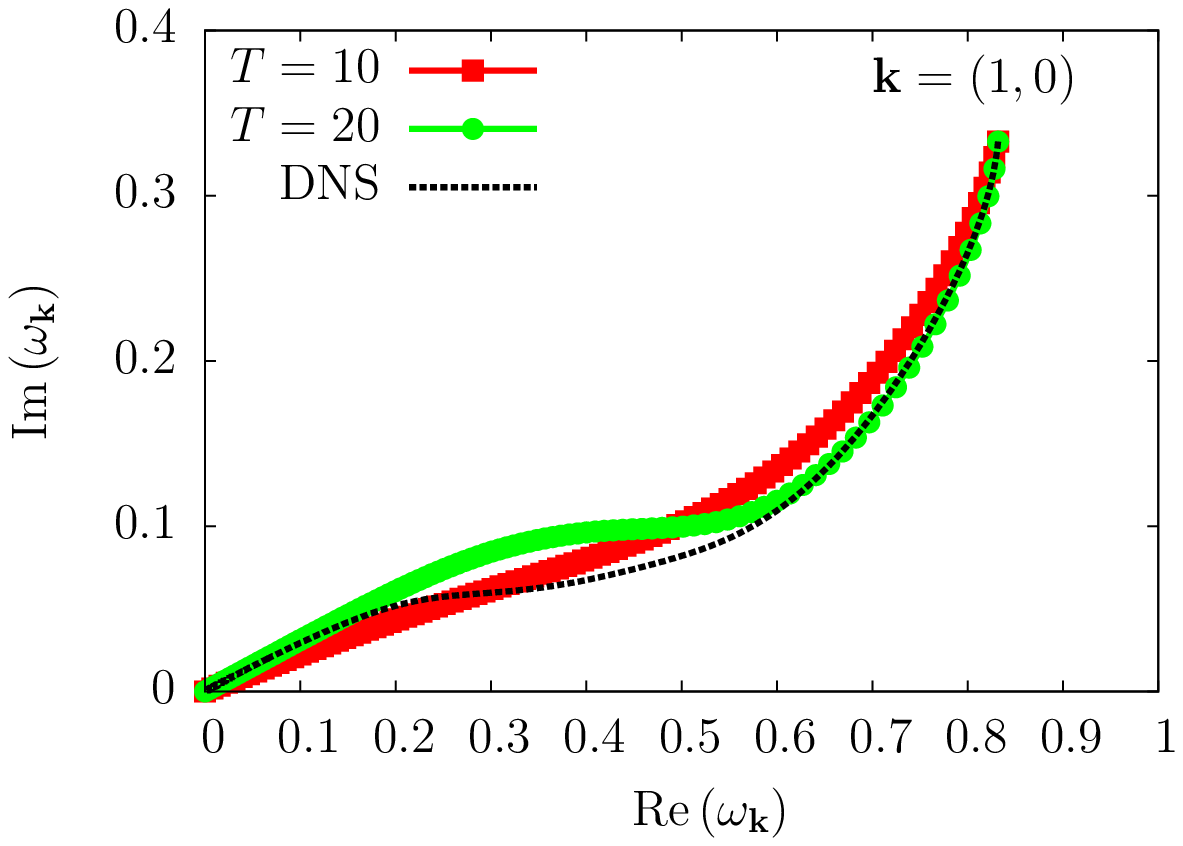}
\hfill
\includegraphics[width = 0.3\columnwidth]{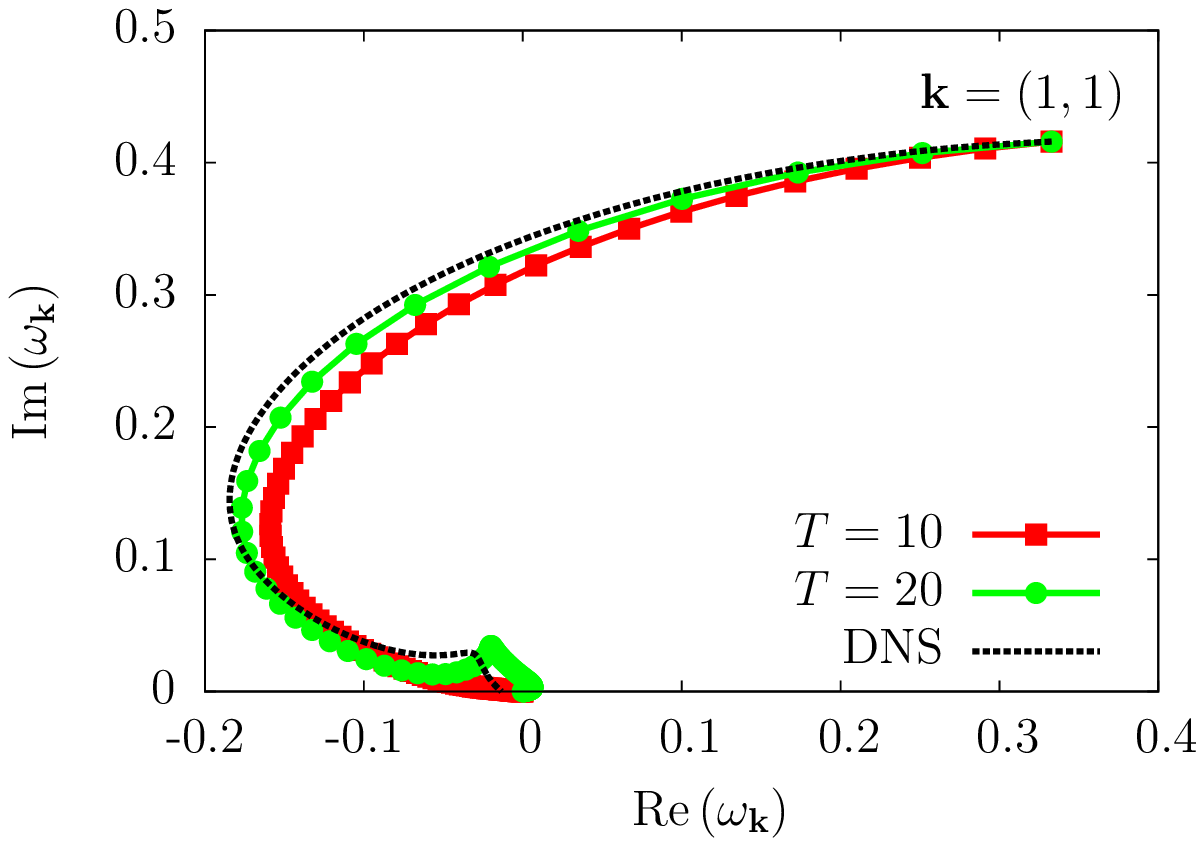}
\hfill
\includegraphics[width = 0.3\columnwidth]{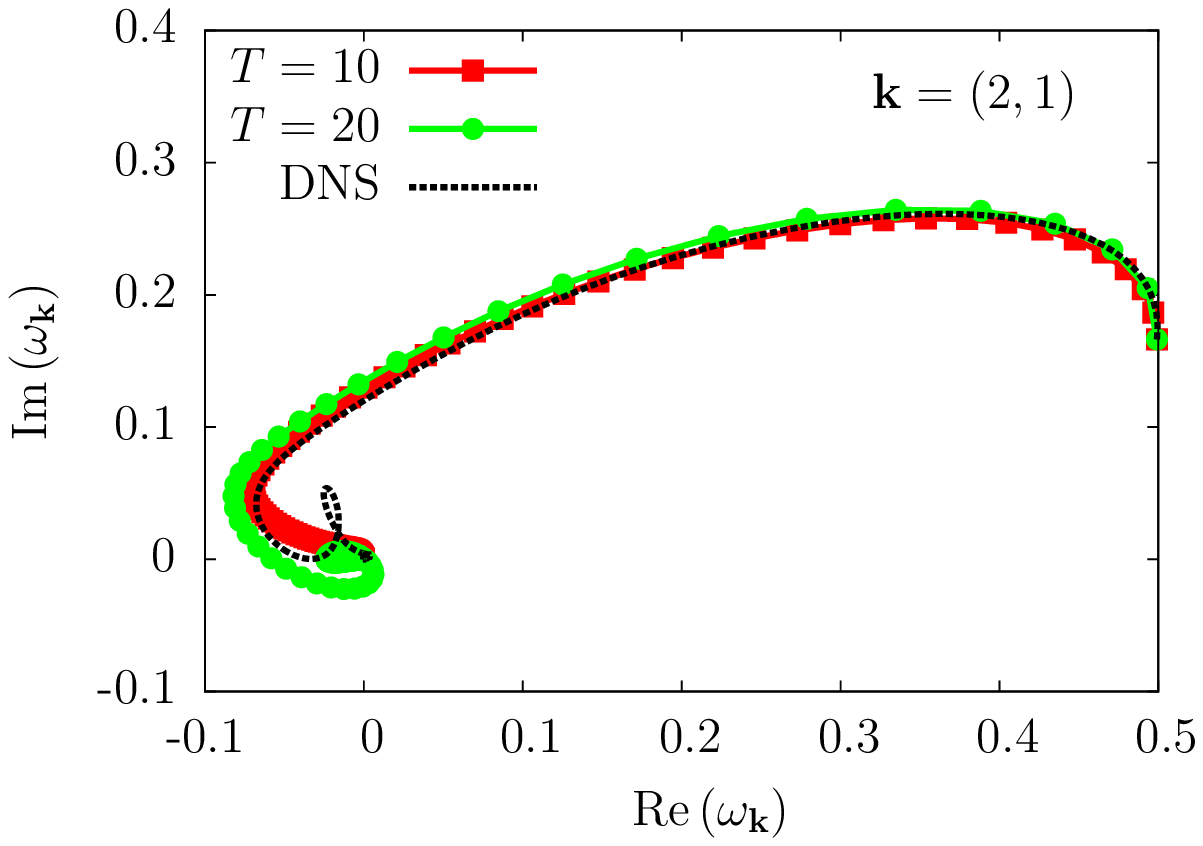}
\hfill
\caption{Complex phase space trajectories of each of the Fourier modes of the final state for  $\bk=(1,0)$ (left), $\bk=(1,1)$ (middle),$\bk=(2,1)$ (right) for transition times $T=10$ and $T=20$.  The black dashed curve corresponds to the $T\to\infty$ limit computed using the relaxation equation~\eqref{eq:relaxation-paths-QG}.\label{fig:traj_decay}}
\end{center}
\end{figure}

We plot the complex phase space trajectories for each Fourier mode in Figure~\ref{fig:traj_decay}.  Again, we observe gradual convergence to the theoretical infinite transition time prediction computed through the relaxation equation.  Notice the complex behaviour of the transition associated to the nonlinear nature of the evolution.  Finally, in Figure~\ref{fig:hamiltonian_decay} we plot the instanton Hamiltonian for the numerical minimum action predictions for the various transition times $T$.  We observe fairly good stationarity of the Hamiltonian across the time evolution for transition times $T$.  Notice that the value of the Hamiltonian decreases with increasing transition time $T$. We expect that the value of the instanton Hamiltonian should decrease with increasing transition time $T$.

\begin{figure}[ht!]
\begin{center}
\includegraphics[width = 0.4\columnwidth]{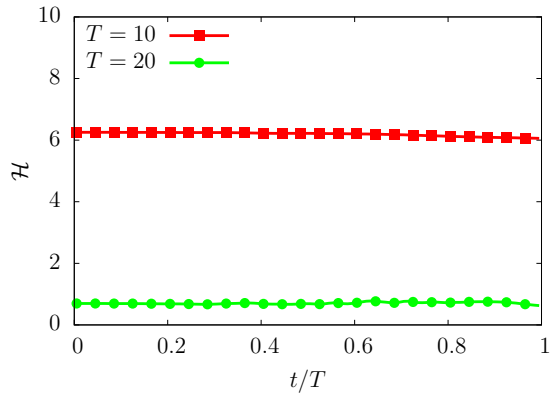}
\caption{Plot of the instanton Hamiltonian for the numerical minimization predictions from the minimum action method for transition times $T=10$ and $T=20$.\label{fig:hamiltonian_decay}}
\end{center}
\end{figure}

Through this example, we have verified the numerical predictions for the most probable rare transition from zero to an arbitrary state using the minimum action method agrees with the theoretical prediction made using the equilibrium hypothesis of section~\ref{sec:equilibrium}.  Through this, we have also independently confirmed that the predictions of rare transitions in equilibrium cases can be predicted through relaxation dynamics of a corresponding dual system.

%We chose our final state to be defined as $\omega_T =[\cos(x)-(2/5)\sin(x) +  (4/5)\cos(y)-(3/5)\sin(y)+ \cos(x+y)-\sin(x+y)+(4/5)\cos(2x+y)-(6/5)\sin(2x+y)+(6/5)\cos(3y)-(7/5)\sin(3y)]/E$, where $E$ being a constant that normalizes the state to unit energy density.  We set $T=10$, with $\alpha=1\time 10^{-1}$.  We consider $N_x=N_y=16$ Fourier modes in each direction and a time resolution of $N_t=1000$.

\subsection{A non-equilibrium geophysical example: A rare transition between two zonal flow states}

In this subsection we will consider a more general example which is of huge interest to the geophysical community. Namely, a rare transition between two zonal jet configurations for the barotropic quasi-geostrophic equations with topography and forced by statistically homogeneous noise.  This example does not verify the equilibrium hypothesis of section~\ref{sec:equilibrium}.

The importance of this example is based upon its practicality.  The rare transition between two zonal flow states is something that arises in Nature, for instance in ocean currents and atmospheric dynamics.  Moreover, the existence of multiple zonal jet attractors in geophysical models has also recently been observed~\cite{constantinou_emergence_2013}, meaning that rare transitions between then in the presence of stochastic fluctuations is an important and viable problem.

Mathematically, the problem is an intriguing one as the set of all zonal states in the periodic barotropic quasi-geostrophic equations ($q(\bx,t)\equiv q(y,t)$) forms a vector space of steady state solutions of the dynamics where the condition $\bv\cdot\nabla q=0$ is always satisfied. By considering the transition between two zonal flow states, there always exists a critical point of the action (a solution of the Euler-Lagrange equations) that remains in the vector space of zonal flows, as long as the noise is non-degenerate in the zonal direction. As discussed below, this zonal critical point of the action verifies simple equations enabling us to make analytical predictions. We stress however, that it is not granted that this zonal critical point is an action minimizer or even a local action minimizer. In this section, for a specific example, we will use the action minimization algorithm to check that this zonal rare transition is actually a local minimizer.

First, let us begin by investigating the theoretical problem. Consider the rare transition between two generic zonal flows (without loss of generality we assume the zonal direction to be $x$), e.g. $q(\bx,0)=q_0(y)$ and $q(\bx,T)=q_T(y)$ with topography only varying across $y$: $h(\bx)=h(y)$. As mentioned previously, we study the transitions between two zonal flows that occur through other zonal states.  Then, the equations for the rare transition are:
\begin{subequations}\label{eq:ansatz_zonal}
\begin{align}
q^*(y,t) = \gamma(t)&\left[q_T(y) - q_0(y) \right] + q_0(y),\\
 \gamma(0)=0,& \quad\gamma(T)=1. 
\end{align}
\end{subequations}
To find the structure of the parameterized path $\gamma(t)$, we insert ansatz~\eqref{eq:ansatz_zonal} into the instanton equations~\eqref{eq:Euler-Lagrange}. Due to the ansatz requiring the transition to remain through zonal steady states, all nonlinear terms identically vanish in the instanton equations.  Then these Euler-Lagrange equations simplify to
\begin{subequations}\label{eq:EL_zonal}
\begin{align}
\frac{\partial^2 \omega^*(y,t)}{\partial t^2}   &=\left[\alpha  + \nu\left(-\frac{\partial^2}{\partial y^2}\right)^{n}\right]^2\omega^*(y,t),\\
\omega(y,0)=&q_0(y) - h(y), \quad \omega(y,T) = q_T(y) - h(y),
\end{align}
\end{subequations}
where $\omega^*(t)=\gamma(t)\left[q_T(y)-q_0(y)\right]+q_0(y)-h(y)$.  The reason for expressing~\eqref{eq:EL_zonal} in terms of the vorticity $\omega$ and not the potential vorticity $q$ is that the topography $h$ only appears in the definition of the boundary states and not the equation of motion itself.  Then, we can straightforwardly solve~\eqref{eq:EL_zonal} for each Fourier amplitude as the system is linear in $\omega$. Subsequently, writing the solution in terms of the Fourier amplitudes for $\omega$ is more transparent
\begin{subequations}\label{eq:instanton_zonal}
\begin{equation}
\omega^*(y,t) = \sum_{k_y}\, \omega_{k_y}^*(t)\be_{k_y}(y),
\end{equation}
with
\begin{equation}
{\omega_{k_y}^*}(t) = \frac{1}{\sinh(\beta_{k_y}T)}\left[\sinh\left(\beta_{k_y}[T-t]\right)\omega_{k_y}(0) +\sinh\left(\beta_{k_y}t\right)\omega_{k_y}(T)\right],
\end{equation}
\end{subequations}
where $\be_{k_y} = \exp\left(ik_y y\right)/L_y^{1/2}$. We have represented the transition in terms of $\omega^*_{k_y}(t)=q^*_{k_y}(t)-h_{k_y}$, where $h_{k_y}$ are Fourier coefficients of the topography $h(y)=\sum_{k_y} h_{k_y}\be_{k_y}$, and $\beta_{k_y}= \alpha + \nu k_y^{2n}$.  Solution~\eqref{eq:instanton_zonal} can be transformed back into the solution for the potential vorticity using $q^*_{k_y}(t)=\omega^*_{k_y}(t)-h_{k_y}$.  What should be noticed is that the transition~\eqref{eq:instanton_zonal} is reminiscent of the over-damped solution presented in subsection~\ref{sub:overdamp}.  This is because the zonal-zonal transition occurs through the vector space of zonal flows and the nonlinearity vanishes throughout the transition.  Therefore, one can think of zonal-zonal transition as the same as the over-damped solution, or in terms of an Ornstein-Uhlenbeck process with the transition exponentially diffusing across steady states.  What is also interesting, is that the solution does not depend on the type of topography, as long as it is defined along $y$ only.

The explicit expression for the Lagrangian for trajectory~\eqref{eq:instanton_zonal} is given by

\begin{align}\label{eq:lag_zonal}
\mathcal{L}\left[q^*, \frac{\partial q^*}{\partial t} \right] =& \frac{1}{2} \int_{\mathcal{D}} \int_{\mathcal{D}}\left[\frac{\partial q^*}{\partial t} + \alpha \omega^* + \nu\left(-\frac{\partial^2}{\partial y}\right)^n \omega^* \right](y)\nonumber
\\&\times C_z^{-1}(y-y') \left[\frac{\partial q^*}{\partial t} + \alpha \omega^* + \nu\left(-\frac{\partial^2}{\partial y'}\right)^n \omega^* \right](y') \, dy \, dy'\nonumber\\
=&\frac{1}{2}\sum_{k_y} \frac{\beta_{k_y}\exp\left(2\beta_{k_y} t\right)}{|f_{k_y}|^2\sinh^2\left(\beta_{k_y}T \right)}\left| \omega_{k_y}(T)-\omega_{k_y}(0)\exp\left(\beta_{k_y} T\right)\right|^2,
\end{align}
where $C_z$ is the zonal part of the noise correlation function defined by~\eqref{eq:zonal_noise}.
% % include numerics % % % % % % % % % % % % % % % % % % % % % % % % % %
To perform the numerical minimization, we select two zonal flow states given by $\omega_0=\left[\cos(y)-(2/5)\sin(y)+(4/5)\cos(3y)-(3/5)\sin(3y)+2\cos(4y)\right]/E$ and $\omega_T=\left[\cos(y)-\sin(y)-(3/2)\sin(2y)+(4/5)\cos(3y)-(4/5)\sin(3y)\right]/E$, where $E$ is the appropriate normalization constant.  Both of these states are displayed in Figure~\ref{fig:w_zonal}.  We use a linear friction coefficient of $\alpha=1\times 10^{-1}$ and normal viscosity ($n=1$) with coefficient $\nu=5\times 10^{-2}$. We consider a transition occurring over a time of $T=10$ with a temporal resolution of $N_t=200$ grid points.  Our Fourier resolution is $N_x=N_y=16$ on a periodic square domain of size $L_x=L_y=2\pi$.  For the noise, our only conditions are that it is homogeneous and non-degenerate.  Therefore, we choose a noise spectrum of the form
\begin{equation}\label{eq:zonal_forcing}
 f_\bk = \frac{1}{Z}\frac{k^2}{k^2_f}\exp\left( - \frac{k^2}{k_f^2}\right)\exp\left(i\frac{\pi}{4}\right),
\end{equation}
with $k_f=3$ and $Z$ being the normalization constant to ensure condition~\eqref{eq:normalization} is satisfied.  The profile of the forcing is shown in Figure~\ref{fig:f_zonal}. As one can observe, the noise is isotropic and peaked at wavenumber $k=3$ with a Gaussian profile around this peak.

\begin{figure}[ht!]
\begin{center}
\hfill
\includegraphics[width = 0.4\columnwidth]{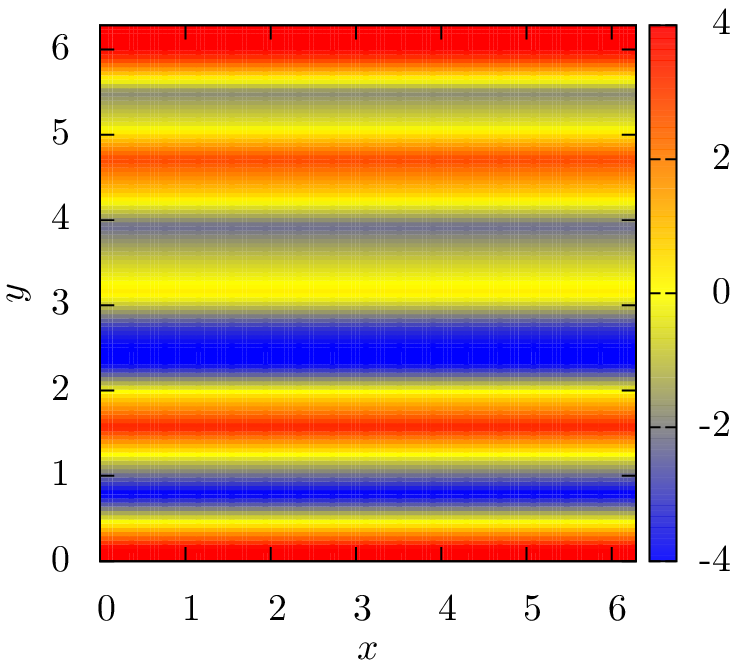}
\hfill
\includegraphics[width = 0.4\columnwidth]{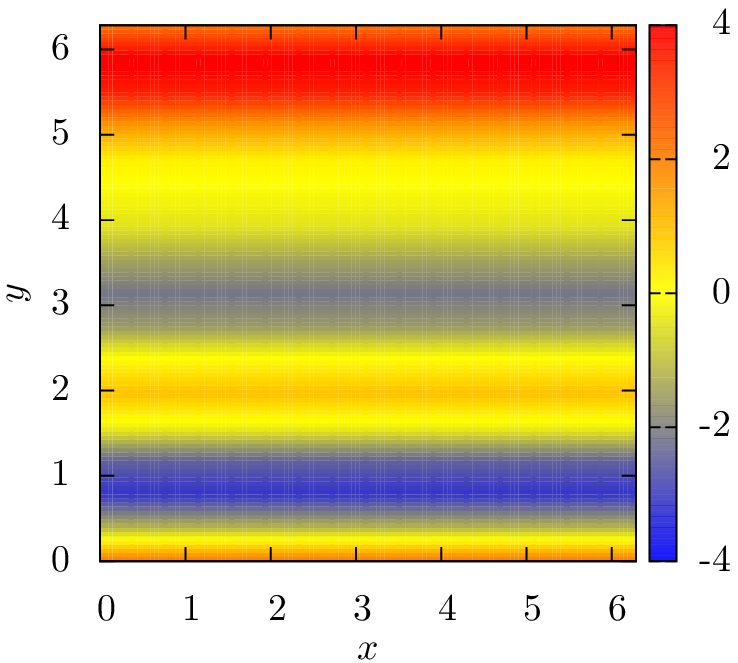}
\hfill
\caption{The vorticity distribution of the initial and final zonal flow states: $\omega_0=\left[\cos(y)-(2/5)\sin(y)+(4/5)\cos(3y)-(3/5)\sin(3y)+2\cos(4y)\right]/E$ and $\omega_T=\left[\cos(y)-\sin(y)-(3/2)\sin(2y)+(4/5)\cos(3y)-(4/5)\sin(3y)\right]/E$ with $E$ the normalization constant to ensure unit energy density in each case. \label{fig:w_zonal}}
\end{center}
\end{figure}

\begin{figure}[ht!]
\begin{center}
\includegraphics[width = 0.5\columnwidth]{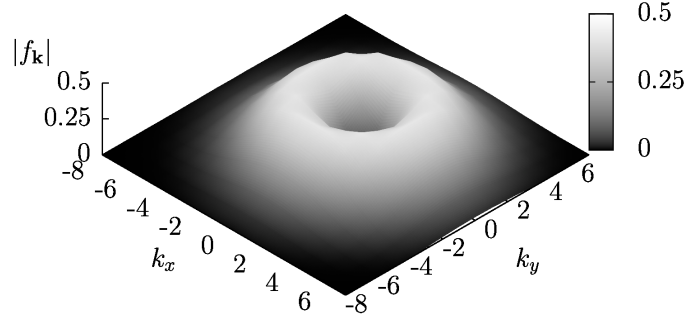}
\caption{Spectral distribution of the noise spectrum $|f_\bk|$, normalized by condition~\eqref{eq:normalization}. \label{fig:f_zonal}}
\end{center}
\end{figure}

In Figure~\ref{fig:zonal_trans} we plot the time evolution of the zonal flow transition from the initial to the final states in terms of an Hovm\"{o}ller diagram. A more illustrative comparison to the theory can viewed in Figure~\ref{fig:zonal_num} which shows the time evolution of the zonal Fourier amplitudes of the numerically predicted transition (left) and also the transition of each mode in the complex phase space (right).  Overlaid in both of the plots in the black dashed curves are the theoretically predicted transition paths from Eq.~\eqref{eq:instanton_zonal}.  We observe excellent agreement between the numerical data and theory, indicating that the minimum action method has located a local action minimizer of the action that coincide with the theoretical zonal critical point.

As additional checks, we present the time evolution of the Lagrangian~\eqref{eq:lag_zonal} and instanton Hamiltonian~\eqref{eq:instantonHam} in Figure~\ref{fig:lag_ham_zonal} with the theoretical predictions overlaid in black.  Again, we observe that the theory agrees with the numerical data.  The Lagrangian indicates that most of the effort is in pushing the transition to the final state near the end of the transition. The constant value of the Hamiltonian throughout the transition is another indicator that the minimum action method has found a local minimum.  

As can be observed from the expression of the Lagrangian for this setup~\eqref{eq:lag_zonal}, the amplitude of the noise plays an essential role in determining which Fourier modes contribute to the Lagrangian and hence the action. An important remark in this example is that the noise correlation will not have a direct effect on the shape of the transition, this being due to the nonlinear terms vanishing, but will be essential for determining the specific value of the  action corresponding to the rare transition.  This value will measure the rarity of the transition and its likelihood of it being observed. 

\begin{figure}[ht!]
\begin{center}
\includegraphics[width = 0.5\columnwidth]{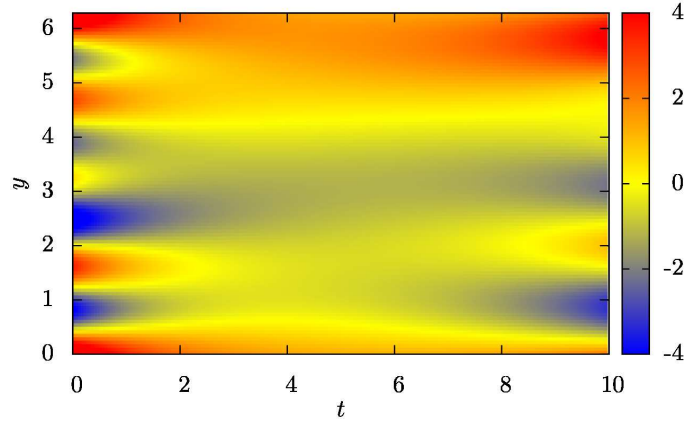}
\caption{Hovm\"{o}ller diagram of the zonal flow transition.\label{fig:zonal_trans}}
\end{center}
\end{figure}

\begin{figure}[ht!]
\begin{center}
\hfill
\includegraphics[width = 0.4\columnwidth]{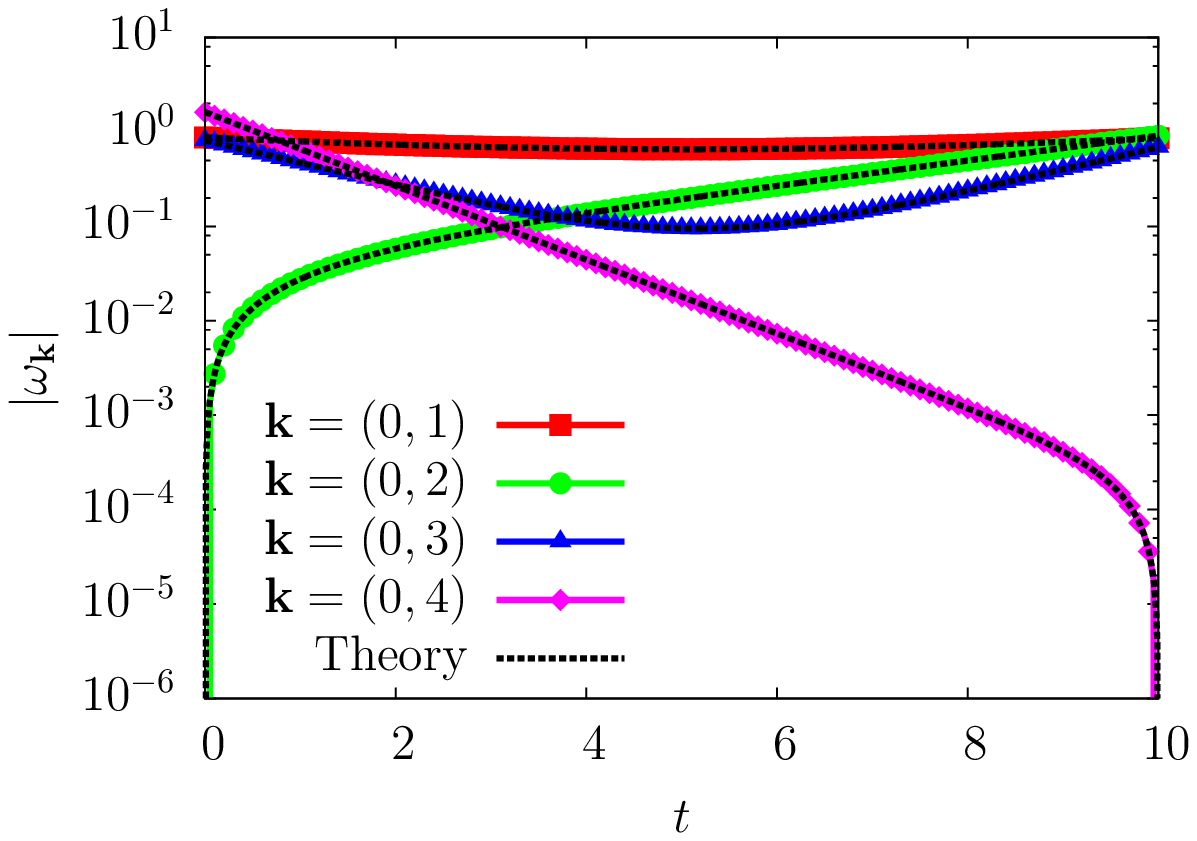}
\hfill
\includegraphics[width = 0.4\columnwidth]{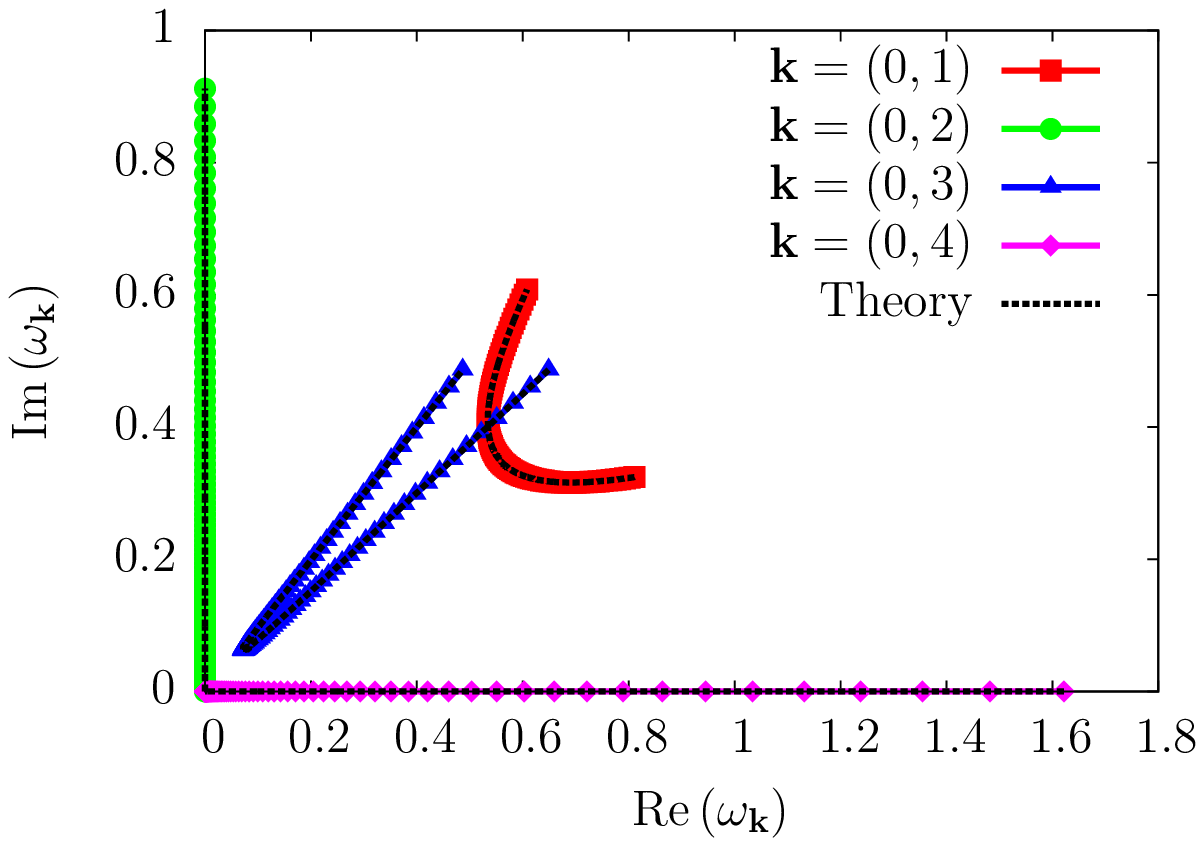}
\hfill
\caption{(Left) Time evolution of the modulus of zonal Fourier harmonics of modes $\bk=(0,1)$, $\bk=(0,2)$, $\bk=(0,3)$ and $\bk=(0,4)$. The black dashed curve overlays the theoretical prediction from~\eqref{eq:instanton_zonal}. (Right) The phase space trajectories of the zonal Fourier harmonics from numerically found transition and the prediction of~\eqref{eq:instanton_zonal}. \label{fig:zonal_num}}
\end{center}
\end{figure}

\begin{figure}[ht!]
\begin{center}
\hfill
\includegraphics[width = 0.4\columnwidth]{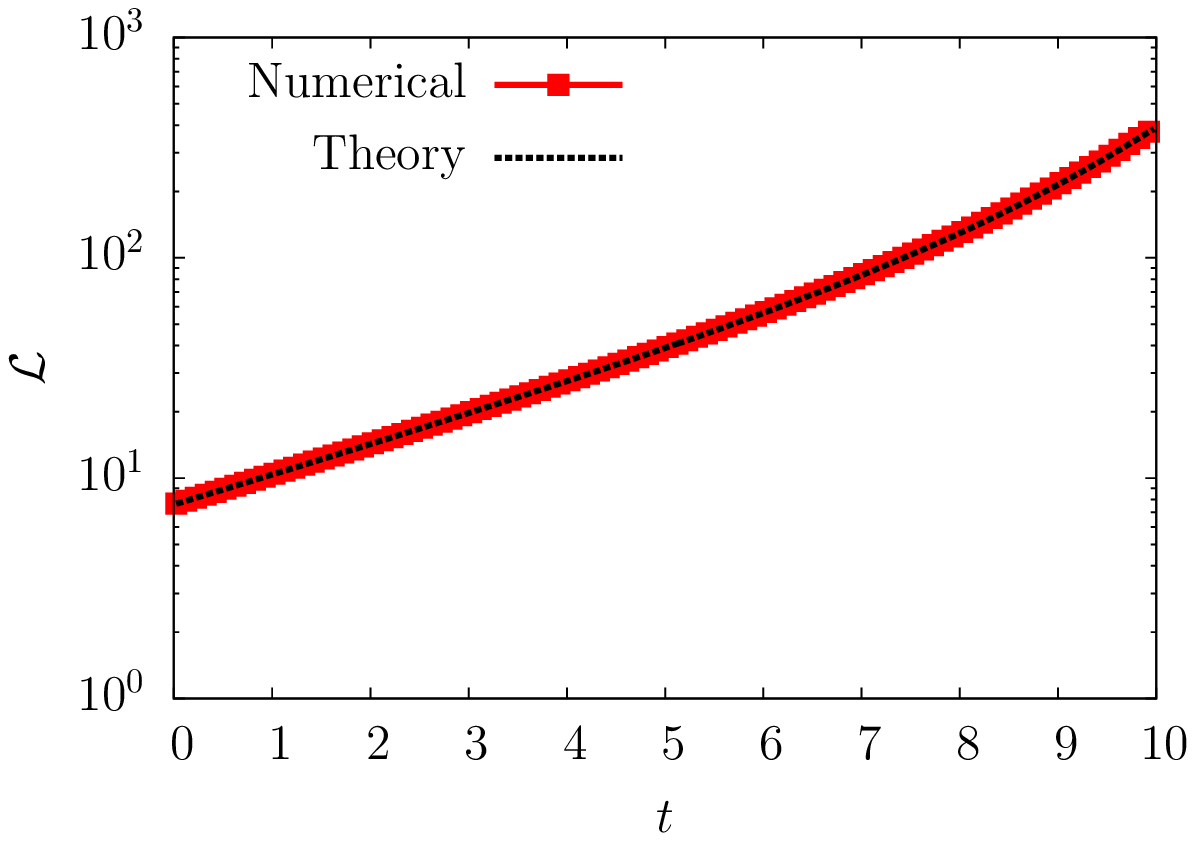}
\hfill
\includegraphics[width = 0.4\columnwidth]{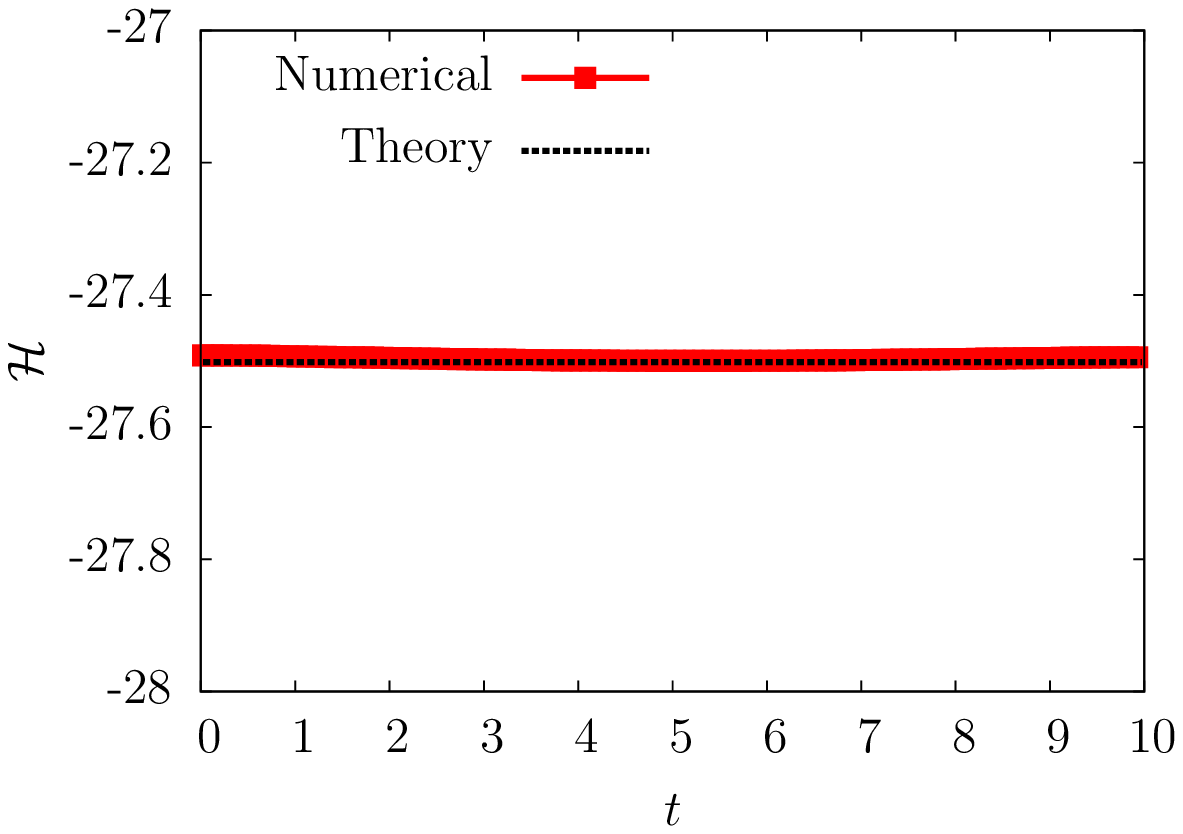}
\hfill
\caption{(Left) Evolution of the Lagrangian~\eqref{eq:lag_zonal} with comparison to the theoretical prediction. (Right) Time evolution of the instanton Hamiltonian~\eqref{eq:instantonHam} with comparison to the theoretical value from transition~\eqref{eq:instanton_zonal}.\label{fig:lag_ham_zonal}}
\end{center}
\end{figure}

% % % % % % % % % % % % % % % % % % % % % % % % % % % % % % % % % % % % %

The example above illustrates how one expects a rare transition between two zonal flow states will occur.  We predict that the transition will remain through the vector space of zonal flows with the structure of the transition being independent on the non-degenerate noise correlation.  A natural next step would be to try and observe such a transition in direct numerical simulations of the quasi-geostrophic equations in regimes like the ones observed in~\cite{constantinou_emergence_2013} where multiple zonal jet have been observed as dynamical attractors. 

The above result on the zonal-zonal transition can be generalized in the context of a rare transitions between two steady states that are formed from eigenfunctions of the Laplacian operator $\Delta$ with the same eigenvalue $\lambda$ where $-\Delta q =\lambda q$. This is because the set of states constructed by eigenfunctions of the Laplacian also form a vector space of steady state solutions with $\bv\cdot\nabla q=0$. Consequently, if both states, initial and final, are constructed with the same sets of eigenfunctions with identical eigenvalues then we expect a similar result as above.

\section{Conclusions}
\label{sec:conclusions}

We have adapted a numerical optimization algorithm called the minimum action method and applied it to a simple model of two-dimensional geophysical turbulence.  We have shown, using specific examples, that such an algorithm can be used to compute the most probable rare transitions between two states in cases of bistability in turbulent systems.  Using equilibrium theory derived in~\cite{bouchet_langevin_2014}, we showed how the numerically predicted transition agreed with those computed through the relaxation equations of the corresponding dual system when the equilibrium hypothesis holds.  Furthermore, we considered a more general problem of computing the most probable rare transition between two different zonal flow configurations where the equilibrium hypothesis does not hold; an important example of relevance to geophysics.

The minimum action method is a viable way to compute rare events in simple turbulent models.  It is straightforward to extend this method to more complex turbulent models such as magneto-hydrodynamics where rare transitions between different magnetic field polarizations can be observed~\cite{berhanu_magnetic_2007}.  Moreover, natural extensions to the algorithm proposed here could be of benefit, such as arc length parameterization of time~\cite{vanden-eijnden_geometric_2008}, adaptive discretization~\cite{zhou_adaptive_2008}, or parallelization~\cite{wan_adaptive_2011}.

Clearly, the next step in this approach is to compare the action minimizers with observed transitions in both experiments and direct numerical simulations.  Indeed, this is one of the current directions of future work. Beside this direct comparison, many work is still to be done both at the theoretical and at the practical level in order to actually assess when minimum action methods alone will be enough to describe rare transitions. This is certainly true when we are in a Freidlin-Wentzell regime, as discussed in subsection~\ref{subsec:LDtheory}, however for most turbulence models there is no clear criteria developed yet to assess when a rare transition in a turbulent flow is actually in the Freidlin-Wentzell regime. This is an important question that should be addressed both from a theoretical and an empirical point of view.

This work is a step in a long term program that is aimed at developing the tools to compute rare transitions and their probabilities in complex turbulent flows. Our ultimate aim is to be able to make these computations for models that are relevant for climate dynamics. Of course much is still to be achieved in this direction before climate applications can be really considered. However it is important to stress that no approach currently allows to reliably compute rare transition in climate problems. 

\section*{Acknowledgments}

The research leading to these results has received funding from the European Research Council under the European Union’s Seventh Framework Programme (FP7/2007-2013 Grant Agreement no. 616811) (F. Bouchet). The research has also been supported through the ANR program STATOCEAN (ANR-09-SYSC-014) (J. Laurie) and the PSMN (P\^{o}le Scientifique de Mod\'{e}lisation Num\'{e}rique) computing center of ENS de Lyon.

\section*{References}
\bibliographystyle{IEEEtran}

\bibliography{bib_jason2}

\end{document}